\documentclass[manuscript]{aastex}
\usepackage{graphicx}
\usepackage{verbatim}
\usepackage{morefloats}
\usepackage{subfigure}
\usepackage{color}
\usepackage{enumerate}
\usepackage{pdflscape}
\usepackage{amsmath,amssymb}

\def\bd{\begin{displaymath}}
\def\be{\begin{equation}}
\def\ed{\end{displaymath}}
\usepackage[normalem]{ulem} 
\usepackage{soul} 
\def\ee{\end{equation}}

\shorttitle{Constraints from Astronomical Observations}
\shortauthors{Burch \& Cowsik}

\begin{document}

\title{Properties of Galactic Dark Matter:  Constraints from Astronomical Observations}
\author{B. Burch\altaffilmark{1,2} and R. Cowsik\altaffilmark{1}}
\affil{Physics Department and McDonnell Center for the Space Sciences, Washington University, St. Louis, MO, 63130}

\altaffiltext{1}{Campus Box 1105, 1 Brookings Drive, St. Louis, MO 63130}
\altaffiltext{2}{bburch@physics.wustl.edu}

\begin{abstract}
The distributions of normal matter and of dark matter in the Galaxy are coupled to each other as they both move in the common gravitational potential. In order to fully exploit this interplay and to derive the various properties of dark matter relevant to their direct and indirect detection, we have comprehensively reviewed the astronomical observations of the spatial and velocity distributions of the components of normal matter. We then postulate that the phase-space distribution of dark matter follows a lowered-isothermal form and self-consistently solve  Poisson's equation to construct several models for the spatial and velocity distributions of dark matter. In this paper, we compute the total gravitational potential of the normal and dark matter components and investigate their consistency with current observations of the rotation curve of the Galaxy and of the spatial and velocity distributions of blue horizontal-branch and blue straggler stars. Even with this demand of consistency, a large number of models with a range of parameters characterizing the dark matter distribution remain. We find the best choice of parameters, within the range of allowed values, for the surface density of the disk 55 M$_\odot$ pc$^{-2}$, are the following: the dark matter density at the Galactic center $\rho_{DM,c}\approx 100-250$ GeV cm$^{-3}$, the local dark matter density $\rho_{DM}(R_0) \approx 0.56-0.72$ GeV cm$^{-3}$, and the root-mean-speed of dark matter particles $< v_{DM}^{2}(R_0 )> ^{1/2}\approx490-550$ km s$^{-1}$. We also discuss possible astronomical observations that may further limit the range of the allowed models. The predictions of the allowed models for direct and indirect detection will be discussed separately in a companion paper.
\end{abstract}

\keywords{dark matter --- Galaxy: fundamental parameters --- Galaxy: structure}
\maketitle

\section{Introduction}
In the study of dark matter, in its role in the formation of structures in the Universe, and in its direct and indirect detection, the mass and luminosity distribution of the Milky Way Galaxy plays a central role. Such considerations perhaps began with the contributions of \citet{Oort32} to the understanding of the problem in terms of the mass model of the Milky Way and is still relevant. The importance of dark matter, especially weakly interacting relics from the Big Bang \citep{Cowsik72,Cowsik73,Lee77,Kolb}, in the formation and dynamics of galactic systems became well-established in the decades following the early 1970's when detailed and critical reviews were written, and modeling of the luminosity distributions and kinematic probes of the Galactic potentials were undertaken \citep{Schmidt56,Bahcall80,Caldwell81,Rohlfs88}. Since that time, systematic improvements of the mass models have taken place, notably by \citet{Dehnen98}, who were concerned with shedding light on the spheroidal nature of the Galactic halo. To this end, they calculated the Galactic gravitational potential for an axisymmetric mass model and investigated its concordance with astronomical constraints obtained from the observations including terminal velocities in the inner Galaxy, the rotation curve, Oort constants, satellites of the Milky Way, the local surface density of the disc \citep{Kuijken91}, etc. Even though the conceptual basis of the mass models have remained more or less the same, significant sophistication in constraining the parameters of the models, such as Bayesian or Markov Chain Monte-Carlo methods, have led to significant progress \citep{Weber09,McMillan}. With progressively improving capabilities in computation, numerical simulations of galaxy formation provide new mass profiles such as that by \citet{Navarro97}, which have provided further stimulus to addressing issues related to the mass models, including forms that avoid the central cusp, which could have been smoothened by baryonic infall \citep{Nesti13}. The phenomenal sensitivities for the direct detection of the particles of dark matter with instruments placed underground \citep{CDMS,XENON,COUPP,PICASSO,SIMPLE} have brought the focus of these studies to the velocity distribution of the dark matter particles, as this is needed for the proper analysis of the signals detected in these experiments. To this end, several authors have adopted the Eddington formula \citep{BT}, which allows the calculation of the velocity distribution of particles that will self-consistently give rise to a spherically symmetric density distribution  that generates the spherically symmetric gravitational potential. By approximating the gravitational potential of the Galaxy to be spherically symmetric, the distribution function that will yield a chosen spherically symmetric mass model for the dark matter halo can be determined \citep{Catena12,Bhatta13}. It is clear from this brief review that the development of the mass models has progressed systematically and has reached a high level of sophistication.

With these aforementioned developments, mass models have reached an important watershed. To proceed further one may take recourse to the well-established paradigm in many branches of Physics and Astrophysics and try to specify the form of the phase-space distribution for all the components of the Galaxy, including the dark matter. If this can be achieved, all the quantities of astrophysical interest may be calculated from these distributions. However, this is a far too challenging task at this stage and one must approach the problem in steps. In this paper, we have adopted an alternative approach \citep{Cowsik96} to the modeling of the Galaxy: we start with a functional form for the phase-space distribution of dark matter, which is motivated by physical considerations and has parameters that have a direct bearing on the problem at hand. We then take recourse to astronomical data to obtain a parameterized description of the mass distribution of baryonic (visible) matter in the Galaxy. We then calculate, \emph{self-consistently}, how dark matter with the specified phase-space distribution will distribute itself in the gravitational potential generated by the dark matter itself and that generated by the visible matter in the Galaxy. Then, the same set of kinematic and other constraints derived from astronomical observations used in constraining the mass models is used to derive the parameters of the phase-space distribution. This approach has not been used extensively, but the earlier studies have indicated some benefits to having a physically motivated distribution function on hand to address the problems related to direct and indirect detection of dark matter. Even this restricted approach already allows us to estimate, for example, the allowed range of the dark matter densities in the central regions of the Galaxy, a parameter difficult to obtain from other methods which rely exclusively on the kinematic indicators like dispersion of the bulge stars and rotation curves in the central regions of the Galaxy, which are dominated by the visible matter distributions in those regions. The phase-space distribution is the same throughout the Galaxy and as such, can efficiently probe the properties of the system in the entire region accessed by the dynamical trajectories.

As stated above, though the precise phase-space distribution of Galactic dark matter is still not well understood, much effort has been invested in understanding the distribution and dynamics of dark matter in the Milky Way for the purposes of understanding the overall dynamics and structure of our Galaxy as well as for the planning and interpretation of their direct and indirect detection. Most studies assume the framework of the Standard Halo Model of dark matter, which envisages the Milky Way as embedded in an isothermal dark matter halo described by a Maxwell-Boltzmann phase-space distribution (often without ensuring self-consistency) with a local dark matter density of 0.3 GeV cm$^{-3}$ and a velocity dispersion of 270 km s$^{-1}$, and the radial distribution of its density is truncated at the ``virial radius'' to keep the mass of the halo and the escape velocity finite. However, it is well understood that a Maxwellian distribution is not appropriate for the description of Galactic dark matter \citep{Kuhlen10}, as it leads to infinite spatial extent and a total mass for the system that indefinitely increases linearly at large distances. Also, recent attempts to estimate the local dark matter density leads to a wide range of values $\sim0.2-0.6$ GeV cm$^{-3}$: $\sim0.39$ GeV cm$^{-3}$ \citep{Catena10}, 0.2-0.4 GeV cm$^{-3}$ \citep{Weber09}, $0.40\pm0.04$ GeV cm$^{-3}$ \citep{McMillan}, $0.43\pm0.11\pm0.10$ GeV cm$^{-3}$ \citep{Salucci10}, 0.3$\pm0.1$ GeV cm$^{-3}$ \citep{Bovy12}. In this paper, we carry out a self-consistent calculation of the spatial distribution of dark matter, including the effects of the background gravitational potential generated by the baryonic matter in the Galaxy. Comparison of the predictions of the self-consistent model with various kinematic observables constrain the parameters characterizing the phase-space distribution of Galactic dark matter. This allows us to derive the density distribution and other properties pertaining to Galactic dark matter particles in a self-consistent way.

We follow the basic strategy for probing the phase-space distribution (PSD) of dark matter developed by \citet{Cowsik96}. The distributions of dark matter and visible matter are coupled to each other as they are both influenced by the common gravitational potential of the Galaxy, to which each component makes its own contribution. Thus, when we have at hand the density distribution of visible matter, determined by astronomical observations, it is a straight-forward matter to calculate the gravitational potential generated by visible matter. Then, for any assumed functional form for the PSD of dark matter, with a given a set of parameters, we can solve the non-linear Poisson equation to determine the potential contributed by dark matter. Thus, having determined the total potential of the Galaxy, we may then use astronomical probes like the rotation curve of the Galaxy and the velocity and spatial distribution of stars to determine the values of the parameters characterizing the PSD function of dark matter that provide good fit to these observations.

The accuracy of determination of the parameters characterizing the phase-space distribution of Galactic dark matter depends primarily on the accuracy and extent of astronomical data. Accordingly, the observations of the distribution of the various visible mass components of the Milky Way are reviewed in detail, and correspondingly, a simple axisymmetric model of the density distribution of the Galactic stars and gas is constructed. The particles of dark matter move in the gravitational potential generated by the visible matter and by their own mass distribution, and their spatial distribution is generally assumed to be cut off at the ``virial radius'', defined as the radius of a sphere which has an average density of $\Delta$ times the critical density. The value of $\Delta$ suggested by different authors ranges from $\Delta=200$ \citep{McMillan} to $\Delta=355$ \citep{Libeskind10} or more. However, our analysis indicates that many of the parameters of the phase-space distribution function of dark matter are only weakly sensitive to this choice. Assuming a lowered-isothermal phase-space distribution for Galactic dark matter, the gravitational potential from dark matter is computed from iterative, self-consistent, solutions of Poisson's equation, following a method adopted by \citet{Cowsik96}.  In this way, many dark matter profiles are generated with a wide range of properties depending on the choice of the parameters specifying their PSD. These models are then compared in turn with the current observations of the Galactic rotation curve and to recent observations of the velocity and spatial distribution of blue horizontal-branch (BHB) and blue straggler (BS) stars in the outer Galaxy. Comparing the dark matter models with the rotation curve and with the BHB/BS distributions separately allows for a wide range of dark matter properties. On the other hand, when we demand that the models fit both these simultaneously for the same choice of parameters, the parameter space narrows and becomes almost exclusively dependent on the value chosen for the surface density and scale lengths of the visible disk of the Galaxy. However, even in the combined analysis of the rotation curve and the stellar distributions, a significant range of parameters for the phase-space distribution of dark matter is still allowed. Within the current estimates of the mass of the Galactic disk, we find that the best estimates for the local dark matter density are $\sim$0.56-0.72 GeV cm$^{-3}$ with the value of the density at the Galactic center ranging from 100-250 GeV cm$^{-3}$. The escape speed from the center of the Galaxy lies in the range $\sim 940 - 980$ km s$^{-1}$, and the root-mean-square speed of dark matter particles in the Solar neighborhood is found to be $\sim$490-550 km s$^{-1}$ for the most favored models.

\section{The Visible Matter Distribution}
In order to derive the Galactic dark matter distribution, the distribution of visible matter in the Galaxy must be well understood, as it is their interplay that allows us to probe the dark matter. We have completed an extensive survey of the current observations of the distributions of stars and gas in the Milky Way and constructed a simple axisymmetric model of  the Galaxy that agrees with current data. The known mass distribution of the Galaxy derives contributions from the central black hole, central bulge, disk, and stellar halo populations. In this analysis, the stellar halo is neglected because its contribution to the visible matter density is on the order of 0.1\% at $R_0$, the distance from the Sun to the Galactic center \citep{Helmi}, and it is therefore expected to contribute negligibly to the dynamics of the Galaxy. Likewise, the black hole and nuclear bulge, whilst playing important  roles very close to the Galactic center,  contribute negligibly to the overall dynamics of the Galaxy in the regions of interest. For the purposes of computing the gravitational potential in this paper, the visible Galaxy consists of a central bulge and a thin and a thick exponential disk. There are several ways to model these components, and here, models that most closely fit the  observations of the stars and gas are adopted. 

\subsection{The Disk}
The mass density of both the thin (tn) and the thick (tk) disk components are typically modeled as double-exponential functions with early evidence for their validity given by \citet{Freeman}. Following the notation used in many recent analyses \citep{Cabrera,McMillan}, the densities for the separate components may be written as, 
\begin{eqnarray}
\rho_{tn}(r,z)&=&\frac{\Sigma_{tn,c}}{2z_{tn}}e^{-|z|/z_{tn}}e^{-r/r_{tn}}\label{dtn},\\
\rho_{tk}(r,z)&=&\frac{\Sigma_{tk,c}}{2z_{tk}}e^{-|z|/z_{tk}}e^{-r/r_{tk}}\label{dtk},\\
\rho_d&=&\rho_{tn}+\rho_{tk}\label{disk},\\
\Sigma_d&=&\Sigma_{tn}+\Sigma{tk},
\end{eqnarray}
in cylindrical coordinates, where $\rho_d$ is the total density of the Galactic disk, $\Sigma_{tk/tn,c}$ are the surface densities near the Galactic center (including both stars and gas), $z_{tk/tn}$ are the scale heights, and $r_{tk/tn}$ are the radial scale lengths for the thick and thin components respectively. The masses of the two components are given by
\be M_{tk/tn}=2\pi \Sigma_{tk/tn,c}r^2_{tk/tn}.\ee 
For the total local surface density at $R_0$ from visible matter, we consider the values $\Sigma_{d,\odot}=$40, 55, and 70 M$_{\odot}$ pc$^{-2}$, which is within the range found in the literature  \citep{Kuijken89b,Kuijken91,Flynn94,Gould96,Korchagin03,Siebert03,Holmberg04,Flynn06,Weber09}. For convenience, the ratio of the thick disk surface density at $R_0$ to the total surface density at $R_0$ is taken to be
\be\frac{\Sigma_{tk,\odot}}{\Sigma_{tk,\odot}+\Sigma_{tn,\odot}}=0.1,\label{sigratio}\ee
a value that is also within the range of the findings of many studies \citep{Ojha01,Reid93,Buser,Spagna,Ng,Larsen,Robin96,Siegel,Carollo}. The choice of the ratio in eq. \ref{sigratio}, within observational constraints, does not have a significant effect on the rotation curve or other dynamical indicators calculated in this paper. 

From the compilation of observations, a complete range in the radial scale length of the thin disk $r_{tn}$=2.0, 2.5, 3.0, and 3.5 kpc was considered, and $r_{tk}$ was taken to  be 3.5 kpc, though there is recent evidence that the radial scale length of the thick disk could be as short as $\sim2$ kpc \citep{Bensby11,Cheng12,Bovy12b}. The scale heights are chosen to be $z_{tn}=350$ pc and $z_{tk}=900$ pc, close to the values determined by  \citet{Juric} and are seen to not have a strong effect on the calculated rotation curve, agreeing with the analysis of \citet{McMillan}. The above parameters yield a combined disk mass in the range $3.57-6.24\times10^{10}$ M$_\odot$. Note, this simple disk model does not include some of the finer structures of the disk such as the spiral arms or the warp included in other analyses \citep{Sofue08} since an axisymmetric model is adequate to describe the overall dynamics and accords computational simplicity. 

The disk potential has the form \citep{Kuijken89b,Mo10}
\begin{eqnarray} \Phi_d(r,z)&=&-2\pi G\Bigg(\Sigma_{tn,c}r_{tn}^2\int_0^\infty dk\frac{J_0(kr)}{[1+(k r_{tn})^2
]^{3/2}}\frac{e^{-k|z|}-(k z_{tn})e^{-|z|/z_{tn}}}{1-(k z_{tn})^2}\nonumber\\
&&+\Sigma_{tk,c}r_{tk}^2\int_0^\infty dk\frac{J_0(kr)}{[1+(k r_{tk})^2
]^{3/2}}\frac{e^{-k|z|}-(k z_{tk})e^{-|z|/z_{tk}}}{1-(k z_{tk})^2}
\Bigg).
\end{eqnarray}
Note the potential includes integration over Bessel functions, $J_0(kr)$, which slows down numerical calculations of the rotation curve for the different models. We find that using an adaptive quasi-Monte Carlo method in \emph{Mathematica} provides the quickest computation time without sacrificing accuracy.


\subsection{The Bulge}
\begin{figure}[h]
\centering
\includegraphics[width=17cm]{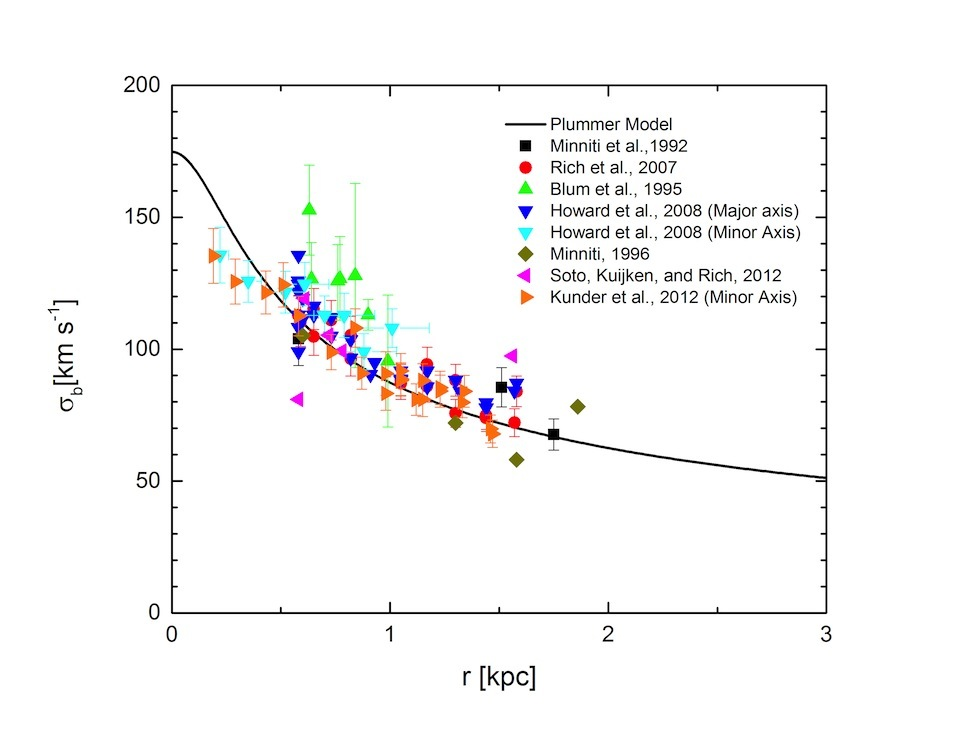}
\caption{The velocity dispersion expected from the bulge (eq. \ref{bdispeq}) is plotted along with K and M giant observations \citep{Minniti92,Rich,Blum95}. The agreement between the observations and the prediction from the Plummer profile implies that the mass distribution of the bulge can be adequately fit by a Plummer profile in the region of interest.}\label{bdisp}
\end{figure}

The  bulge, in cylindrical coordinates, is described by a Plummer density profile of the form
\be \rho_b=\frac{3M_b}{4\pi b^3}\Bigg(1+\frac{r^2+z^2}{b^2}\Bigg)^{-5/2},\label{bulge}\ee
where $M_b$ is the total mass of the bulge and $b$ is the scale radius. To determine $M_b$ and $b$, we assume the dark matter contribution to the dynamics from $\sim0.1-1$ kpc is small (an assumption that implies that visible matter dominates the dynamics in the very central regions of the Galaxy and is justified \emph{post facto}) and subtract the disk contribution from the inner 1 kpc. Noting that the contributions of various components to $v_c^2(r)$ are additive, we subtract the theoretically expected contribution to $v_c^2$(r) from the square of the observed values for $r\lesssim1$ kpc to obtain the exclusive contribution of the bulge. The resulting points are fit with the rotation curve derived from the Plummer density. The bulge parameters are found to be $M_b=1.02\times10^{10}~M_\odot$ and $b=0.258$ kpc. The potential of the Plummer model has the simple analytical form
\be \Phi_b=-\frac{G M_b}{(r^2+z^2+b^2)^{1/2}}\ee
and a velocity dispersion \citep{Dejonghe87} given by 
\be \sigma_{b}^2=\frac{G M_b}{6(r^2+z^2+b^2)^{1/2}}. \label{bdispeq}\ee
We show the velocity dispersion in Fig. \ref{bdisp}, along with measurements of K and M bulge giants \citep{Minniti92,Rich,Blum95}. Note that the dispersion expected from the Plummer model agrees well with the observations, implying that this region is adequately described by a bulge-dominated density profile. Including the mass contribution from the disk in the bulge region does not noticeably change the predictions for the velocity dispersion. 

\section{Kinematic Observations: The Rotation Curve and Stellar Velocity Distributions}

Recent observations of stars, interstellar gas, and masers have been able to extend our understanding of the dynamics of the Galaxy beyond the Solar circle as well as improving observations near the Solar location. The density and dynamics of these objects probe the Galactic gravitational potential. The gravitational potential in the Galactic plane as a function of the distance from the Galactic center can be determined from the rotational speed of the Galaxy in the usual way:
\be \frac{v_c^2}{r}=\bigg|\frac{\partial \Phi_{tot}}{\partial r}\bigg|,\ee
where $\Phi_{tot}$ is the total gravitational potential of the Galaxy, obtaining contributions from both its visible and dark matter components. The best observations of the Milky Way's rotation curve now span the range of galactocentric distances $\sim1-20$ kpc. The Galactic potential can be probed up to $\sim90$ kpc by considering the velocity distribution of BHB  and BS stars. Recent compilations of carefully selected BHB stars by \citet{Xue} and BHB and BS stars by \citet{Brown} are used below to constrain the Galactic dark matter distribution.

\subsection{The Rotation Curve}
\begin{figure}[h]
\centering
\includegraphics[width=17cm]{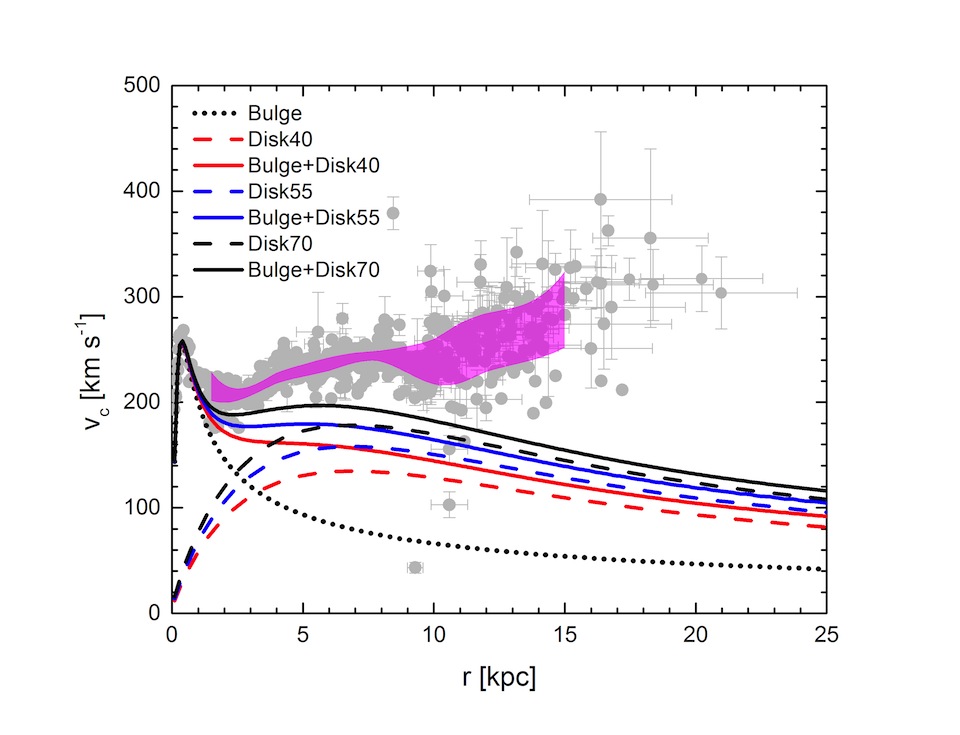}
\caption{Observations of the rotation curve \citep{Burton78,Blitz82,Schneider83,Clemens85,Fich89,Burton93,Turbide93,Honma97,Pont97,Honma07,McClure07,Oh10,Stepanishchev11} are plotted after rescaling all the data using $R_0=8.3$ kpc and $\Theta_0=240$ km s$^{-1}$ and adjusting for the current measurements of the peculiar motion of the Sun. The magenta band indicates there region where 2/3 of the points lie within 1 kpc radial bins. The rotation curve predicted by the visible matter components alone for the three local disk surface densities with $r_{tn}=3.0$ kpc is shown along with the shaded observations.}\label{VisRC}
\end{figure}

The best estimates of the gravitational potential in the Galactic plane within $\sim 10$ kpc of the Galactic center come from measurements of the Galactic rotation speed. We have compiled a large sample of the available observations \citep{Burton78,Blitz82,Schneider83,Clemens85,Fich89,Burton93,Turbide93,Honma97,Pont97,Honma07,McClure07,Oh10,Stepanishchev11}, excluding only those with exceptionally high dispersion in the data \citep{Demers07}.  The rotation curve inside the solar circle is well-determined by observations of HI regions and CO emission associated with HII regions. Outside the solar circle, distances to objects are much more difficult to measure accurately, so the errors in the rotation curve are correspondingly larger. We present a compilation of the data used in our analysis in Fig. \ref{VisRC}, with error bars when available, along with a medial shaded band, which includes $2/3$ of the data points in 1 kpc radial bins and the rotation curve from the visible components of our Galactic mass models with $r_{tn}=3$ kpc. 

The Milky Way's rotation curve is derived from line-of-sight observations of interstellar hydrogen and other material like CO, masers, planetary nebula, and other astrophysical objects. The determination of the rotation curve of the Galaxy depends on the assumed value of $R_0$, the distance from the Galactic center to the Sun and $\Theta_0$, the rotation speed of the Sun about the Galactic center. Recently, maser observations and measurements of stellar orbits near SgrA$^*$ have been able to constrain $R_0=7.2-9$ kpc \citep{Ghez08,Gillessen09,Reid09,Reid08,Brunthaler11}. A summary of some estimates of $R_0$ can be found found in \citet{Avedisova05}. We choose $R_0=8.3$ kpc based on these and other stellar observations from the past decade. The ratio $\Theta_0/R_0$ is well constrained from masers and stellar orbits \citep{Reid09,Brunthaler11,Reid04} and is in the range of $\Theta_0/R_0=28.5-30.3$ km s$^{-1}$ kpc$^{-1}$. We take $\Theta_0/R_0\approx28.9$ km s$^{-1}$ kpc$^{-1}$ such that $\Theta_0=240$ km s$^{-1}$. This value of $\Theta_0$ differs significantly from the IAU standard values (see \citet{McMillan10} and the references therein) of $R_0=8.5$ kpc and $\Theta_0=220$ km s$^{-1}$, which gives $\Theta_0/R_0\approx25.9$ km s$^{-1}$ kpc$^{-1}$ and does not agree with the recent observations of masers and stellar orbits. There are, however, some recent studies which call into question this high value of $\Theta_0$ \citep{Koposov10,Bovy12c}, preferring values closer to 220 km s$^{-1}$.
The result of reanalyzing the observations using the larger value of $\Theta_0$ is that the rotation curve gently rises from $\sim2-15$ kpc instead of remaining flat, and correspondingly, the density of dark matter needed to reconcile the theoretically computed rotation curve with the data increases. We have also corrected all the rotation curve data taking account of the new measurements of the peculiar motion of the Sun ($U_\odot,V_\odot,W_\odot$), where $V_\odot$ has been updated from $\sim5$ km s$^{-1}$ to $\sim11-15$ km s$^{-1}$ \citep{McMillan10,Schonrich10}. We take $(U_\odot,V_\odot,W_\odot)=(11,12,7.5)$ km s$^{-1}$. 

\subsection{Blue Horizontal-Branch Stars}
\citet{Xue} have compiled an extensive list of the line-of-sight velocities of 2401 BHB stars from the Sloan Digital Sky Survey DR6 taking care to ensure their sample is pure and contains accurate data on both the distance and line-of-sight velocity of each star. They use this data to constrain cosmological simulations and estimate the virial mass of the Galaxy as well as derive the rotation curve up to $\sim 60$ kpc. We combine their compilation of 2401 BHB stars within 60 kpc of the Galactic center with their separate compilation of stars at galactocentric distances of $\sim60-90$ kpc.  This combined compilation extends from 5-90 kpc with line-of-sight velocities spanning $\pm350$ km s$^{-1}$. After rescaling the galactocentric distances and line-of-sight velocities for each star to correspond to the values of $R_0$ and $\Theta_0$ chosen above, the observations are divided into eleven radial bins where the positive and negative radial velocities are averaged and divided into 50 km s$^{-1}$ bins. Error bars are shown as $\pm\sqrt{N}$ where $N$ is the average number of stars in each velocity bin.

\citet{Brown} have compiled a sample of 910 BHB and BS stars from the Hypervelocity Star Survey, which contains twice as many stars at $r\geq50$ kpc compared to the compilations by Xue et al., and they derive the velocity dispersion profile of the Milky Way out to $\sim95$ kpc. While the Brown et al. sample claims to be complete in color, magnitude depth, and spatial coverage, there is some ambiguity in distinguishing BHB from BS stars. To achieve the ratio of BHB to BS stars as stated in Brown et al., stars with $f_{BHB}\geq0.6$ were taken to be BHB stars, where $f$ is the likelihood of a candidate star being a BHB star as reported in Table 1 of Brown et al., and any star with $f_{BHB}<0.6$ was considered to be a BS star. This reproduces the  $74\%$ to $26\%$ BHB to BS ratio in Brown et al. As with the stars in the compilation by Xue et al., this sample too was rescaled for our adopted values of $R_0$ and $\Theta_0$. We find that we are unable to reproduce the $R_{BHB}$ and $R_{BS}$ distances in Table 1 of \citet{Brown} for their choice of $R_0$, $\Theta_0$ and ($U_\odot,V_\odot,W_\odot$). We calculate the heliocentric distance for the BHB and BS stars using eq. 2 in Brown et al. and convert it to a galactocentric distance in the usual way (see eq. 4 in \citet{Xue}). The stellar observations are then divided into eight radial bins. The line-of-sight velocities are averaged, and the error bars are determined in the same manner as for the Xue et al. distribution. 

\section{The Galactic Dark Matter Distribution}

The density distribution of dark matter is controlled by both the velocity distribution of the dark matter particles and by the total gravitational potential $\Phi_{tot}$ in which they reside. The total potential receives contributions from the density distributions of both the visible and the dark matter components of the Galaxy:
\be \Phi_{tot}=\Phi_b+\Phi_d+\Phi_{DM}, \ee
where $\Phi_b$, $\Phi_d$, and $\Phi_{DM}$ are the potentials contributed by the bulge, the disk, and the dark matter.  In order to develop a self-consistent model for the dark matter in our Galaxy, we require a dynamical model whose phase-space distribution function represents a collisionless system that can be parameterized by the velocity dispersion of the dark matter, the density at either the Galactic center or at $R_0$, and the size of the dark matter halo. A lowered isothermal (King) distribution is well suited for this and is described in \citet{BT}. Unlike the isothermal sphere, the King distribution has a finite total mass, a non-singular central density, and meets all the requirements of the model. Reasons for choosing the King distribution are also discussed in earlier papers \citep{Cowsik07,Cowsik10}.

The distribution function for the King model is given by
\be f_K(\varepsilon) = \left\{ \begin{array}{ll}
\frac{\rho_1}{\big(2\pi\sigma_{DM}^2\big)^{3/2}}\Big(e^{\varepsilon/\sigma_{DM}^2}-1\Big)& \textrm{for $\varepsilon>0$}\\
0 & \textrm{for $\varepsilon\leq0$}
\end{array} \right.\label{king}\ee
where
\be \varepsilon\equiv\Phi_0-\bigg(\frac{1}{2}v^2+\Phi_{tot}\bigg).\ee
The parameter $\Phi_0$ is the potential at $r=r_t$, the ``virial'' (King) radius of the dark halo, and $\rho_1$ and $\sigma$ are parameters that are related but not equal to the central dark matter density $\rho_{DM,c}$ and the dark matter velocity dispersion.

The dark matter density distribution is readily calculated by integrating the distribution function
\be \rho_{DM}=\int f_K~d^3v\label{intf}\ee
and vanishes at $r=r_t$ where $\varepsilon=0$. Defining the scaled potential $\Psi(r,z)$ as
\be \Psi(r,z)\equiv\Phi_0-\Phi_{tot}(r,z),\ee
the density distribution of dark matter may be written as
\begin{eqnarray}
\rho_{DM}(\Psi(r,z))&=&\frac{\rho_1}{(2\pi\sigma_{DM}^2)^{3/2}}\int^{\sqrt{2\Psi(r,z)}}_0 dv~v^2 \Bigg[\exp\Bigg(\frac{\Psi(r,z)-\frac{1}{2}v^2}{\sigma_{DM}^2}\Bigg)-1\Bigg]\nonumber\\
&=&\rho_1 \Bigg[ e^{\Psi(r,z)/\sigma_{DM}^2} \textrm{erf}\Bigg(\frac{\sqrt{\Psi(r,z)}}{\sigma_{DM}}\Bigg)-\sqrt{\frac{4\Psi(r,z)}{\pi\sigma_{DM}}} \Bigg(1+\frac{2\Psi(r,z)}{3\sigma_{DM}^2}\Bigg)\Bigg].\label{dmdensity}
\end{eqnarray}
Note that $\sqrt{2\Psi(r,z)}$ is just the escape speed from the Galaxy of particles at $(r,z)$.

Keeping in mind that $\rho_{DM}$, which is the source for $\Phi_{DM}$, depends on $\Phi_{DM}$ itself, the dark matter potential must be calculated iteratively by numerically solving the Poisson equation,
\be \nabla^2\Phi_{DM}(r,z)=4\pi G \rho_{DM}(r,z)\label{Poisson},\ee
with $\rho_{DM}(r,z)$ given by eq. \ref{dmdensity}. The procedure that we adopt here is a Legendre  polynomial expansion as described in \citet{Cowsik96}, which is based on the earlier work by \citet{Wilson75} and \citet{Prendergast70}. In this way, we produce the density and gravitational potential for visible matter and dark matter separately as well as for the sum of the two components.

For the above calculations, we formulate the King distribution in terms of $\sigma_{DM}$, $\rho_{DM,c}$, and  $\Psi_0/\sigma_{DM}^2$, where $\Psi(r=0,z=0)=\Psi_0$. The precise choices for $\sigma$, $\rho_{DM,c}$, and $\Psi_0/\sigma_{DM}^2$ are made after solving eq. \ref{Poisson} for a range of values, calculating the corresponding rotation curve (eq. \ref{vctot}), and comparing the results to the observations of the rotation curve and the velocity distribution of BHB and BS stars.

\section{Comparison with Astronomical Observations}
In order to compare model predictions with the observed rotation curve and the observations of spatial and velocity distributions of BHB and BS stars, a large sample of possible dark matter models was generated, with values chosen by hand, encompassing $\rho_{DM,c}=1-1000$ GeV cm$^{-3}$ (We consider 1, 5, 10, 25, 50, 75, 100, 250, 500, and 1000 GeV cm$^{-3}$.), $\sigma_{DM}\approx100-1000$ km s$^{-1}$ (in 5 km s$^{-1}$ intervals), and $\Psi_0/\sigma_{DM}^2\approx1-25$ (in unit intervals) for each of the twelve disks models. For each model, $\Psi(r,z)$, $\rho_{DM}(R_0)$, $<v^2_{DM}(R_0)>^{1/2}$, $v_c(R_0)$, the escape velocity $v_{esc}$, and the total mass of the Galaxy $M_{Galaxy}$ are calculated. Bounds were first put on the parameters of the models by placing the constraint 220 km s$^{-1}\leq v_c(R_o)\leq 260$ km s$^{-1}$. Then, models which agreed with the rotation curve were found within these bounds. 

\subsection{Comparison with the Rotation Curve}
The very discovery of dark matter in the Galaxy and the subsequent determinations of its characteristics rest almost exclusively on the measurements of the rotation curve. In this section, we compare the sequence of models that we have calculated with our compilation of observations shown in Fig. \ref{VisRC} to limit the range of parameters characterizing the phase-space distribution function of dark matter. The theoretical estimates of the rotational velocities are calculated using the expression
\be v_c(r)=\Bigg(r\frac{\partial\Phi_{tot}(r,0)}{\partial r}\Bigg)^{1/2}=\Big(r\frac{\partial}{\partial r}[\Phi_{b}(r,0)+\Phi_{d}(r,0)+\Phi_{DM}(r,0)]\Big)^{1/2}\label{vctot}\ee
for the twelve sets of models characterized by the three choices of $\Sigma_{d,\odot}=$ 40 M$_\odot$ pc$^{-2}$, 55 M$_\odot$ pc$^{-2}$, and 70 M$_\odot$ pc$^{-2}$ and the four choices of $r_{tn}=$ 2.0 kpc, 2.5 kpc, 3 kpc, and 3.5 kpc. We display in Tables \ref{T25}-\ref{T35c} the parameters characterizing the dark matter distribution that fit the rotation curve. We show in Fig. \ref{RotationCurves} a selection of better fitting models for the rotation curve (i.e. pass through the shaded band) for each these sets of model. The corresponding density profiles for these models are shown in Fig. \ref{Densities}. The local and central dark matter density, the escape velocity from the $R_0$ and from the Galactic center, and the root-mean-square velocity at $R_0$ for each of these models are also presented in Tables \ref{T25}-\ref{T35c}.

The following points may be noted after a perusal of the figures and tables:
\begin{enumerate}

\item None of the dark matter models were found to be consistent with the rotation curve for a thin disk scale length of 2.0 kpc. For all choices of PSD parameters, the calculated rotation curves were less than 220 km s$^{-1}$ at $R_0$. To be consistent with the rotation curve, a visible disk with $r_{tn}$=2.0 kpc would have to have a local surface density that falls outside observational constraints. For $r_{tn}=2.5$ kpc, only the lightest disk produced dark matter models consistent with the rotation curve. 

\item All the models presented in Tables \ref{T25}-\ref{T35c} provide good fits to the observed rotation curve within the solar circle ($r<R_0$).

\item The larger the assumed value of the surface density of the disk, larger is the range of allowed parameters of the dark matter distributions.

\item The range in the parameters allowed by all the various models that best fit the rotation curve is:
\begin{enumerate}
\item 872.1 km s$^{-1}$$~\lesssim~ v_{esc}(0)~\lesssim~$983.9 km s$^{-1}$
\item 506.3 km s$^{-1}$ $~\lesssim~ v_{esc}(R_0)~\lesssim~$705.0 km s$^{-1}$
\item 25 GeV cm$^{-3}~\lesssim~\rho_{DM,c}~\lesssim~$500 GeV cm$^{-3}$
\item 0.395 GeV cm$^{-3}~\lesssim~\rho_{DM}(R_0)~\lesssim~$0.760 GeV cm$^{-3}$ 
\item 392.2 km s$^{-1}$$~\lesssim~<v_{DM}^2(R_0)>^{1/2}~\lesssim~$546.1 km s$^{-1}$
\item  $0.61\times10^{12}$ M$_{\odot}~\lesssim~M_{Galaxy}~\lesssim~2.00\times10^{12}$ M$_{\odot}$.
\end{enumerate}

\item Even though we plan to discuss elsewhere the implications of these results for direct and indirect detection of dark matter, we may note here that the range in expected signals is large and the local dark matter density for every model is greater than the standard IAU value of 0.3 GeV cm$^{-3}$. However an assumption of a lower value of $\Theta_0$ would lead to a lower expectation for the dark matter density. The signals for direct detection are directly proportional to $\rho_{DM}(R_0)$ and increase at least linearly with $<v_{DM}^2(R_0)>^{1/2}$, so that the allowed rate may vary at least by $\sim8$, even for detectors with very low threshold. The indirect experiments focus usually on detecting annihilation or decay of dark matter from the central regions of the Galaxy. These rates, proportional to $\rho_{DM}^2(0)$ and $\rho_{DM}(0)$, span a range of 400 and 20 respectively. 

\item To investigate the correlations among the various parameters and derived quantities of the dark matter models given in Tables \ref{T25}-\ref{T35c}, the values of $\sigma_{DM}$, $<v_{DM}^2(R_0)>^{1/2}$, $\rho_{DM}(R_0)$, $v_{esc}(0)$, and $v_{esc}(R_0)$ are plotted as a function of the central dark matter density, $\rho_{DM,c}$, in Figs. \ref{CrossPlots}a-e. Also shown is the dark matter density in the Solar neighborhood as a function of $<v_{DM}^2(R_0)>^{1/2}$ in Fig. \ref{CrossPlots}f. The following observations may be made:

\begin{enumerate}
\item The correlation between the central density $\rho_{DM,c}$ and the value of the $\sigma_{DM}$ parameter is shown in Fig. \ref{CrossPlots}a., and the value of $\sigma_{DM}$  is shown to decrease with increasing $\rho_{DM,c}$. The decrease becomes progressively more gentle, and $\sigma_{DM}$ reaches a values of $\sim195$ km s$^{-1}$ at $\rho_{DM,c}=500$ GeV cm$^{-3}$.

\item The root-mean-square velocities of the dark matter particles near the Solar system, $<v_{DM}^2(R_0)>^{1/2}$, decreases with respect to $\rho_{DM,c}$ as shown in Fig. \ref{CrossPlots}b.

\item Similarly, the dark matter density at $R_0$ decreases gently with respect to $\rho_{DM,c}$ as shown in Fig. \ref{CrossPlots}c.
 
\item The escape velocity from the center of the Galaxy and from the location of the solar system gently decreases beyond $\rho_{DM,c}\sim$200 GeV cm$^{-3}$ for a given set of disk parameters (see Figs. \ref{CrossPlots}d and \ref{CrossPlots}e).

\item The value of $\rho_{DM}(R_0)$ increases statistically with increasing $<v_{DM}^2(R_0)>^{1/2}$ as shown in Fig. \ref{CrossPlots}f, thereby increasing the range of expectation of event rates in direct detection experiments. 
\end{enumerate}

\item After solving Poisson's equation, the potential at every point in the Galaxy is known. It is therefore possible to calculate the force profile in any direction. In Fig. \ref{VerticalForce}, the vertical force profile at the Solar location is shown for a selection of the dark matter models which provide a good fit the observations of the rotation curve. With more extensive astronomical observations, the motions of stars above the Galactic plane at $R_0$ could be used to further constrain the dark matter parameters and the surface density of the disk in the neighborhood of the Sun.

\end{enumerate}

\subsection{Comparison with the Distributions of BHB and BS Stars}
The BHB and BS stars, with their distribution extending up to $\sim90$ kpc, serve as tracers of the gravitational potential of the Galaxy, even though they contribute negligibly to the potential. Since only their radial velocities are well determined, we will first write down their radial distribution function under the assumption that their PSD function  follows the King distribution such as that given in eq. \ref{king}. This is achieved by writing $\Psi(r)=\frac{1}{2}v_{esc}^2(r)$ and integrating the PSD function over the transverse velocities:
\begin{eqnarray}
F_{B}(r,v_r)&=&\frac{\rho_B}{(2\pi\sigma_B^2)^{3/2}}\int_{0}^{v_{esc}^2(r)-v_r^2}\Bigg[\textrm{exp}\Bigg(\frac{v_{esc}^2(r)-v_r^2-v_\perp^2}{2\sigma_B^2} \Bigg)-1 \Bigg]\pi dv_\perp^2\\
&=&\frac{\pi\rho_B}{(2\pi\sigma_B^2)^{3/2}}\Bigg\{2\sigma_B^2\Bigg[\textrm{exp}\Bigg(\frac{v_{esc}^2(r)-v_r^2}{2\sigma_B^2} \Bigg)-1\Bigg]-\big(v_{esc}^2(r)-v_r^2\big)\Bigg\}dv_r\label{BHBdist}.
\end{eqnarray}
The expression given in eq. \ref{BHBdist} is suitable for comparison with the observed distributions. However, the samples of BHB and BS stars \citep{Xue,Brown} appear to be incomplete, at least with regard to radial sampling. In order to assess this, we integrate $F_B(r,v_r)$ over the radial velocities to get the radial distribution of number density, which is given by
\begin{eqnarray}
n_B(r)&=&\int_0^{v_{esc}(r)}F_B(r,v_r)dv_r\\
&=&\frac{\rho_B}{(2\pi\sigma_B^2)^{3/2}}\Bigg[\frac{(2\pi\sigma_B^2)^{3/2}}{2}e^{v_{esc}^2(r)/2\sigma_B^2}\textrm{erf}\Bigg(\frac{v_{esc}(r)}{\sqrt{2\sigma_B^2}}\Bigg)-\frac{2\pi}{3}v_{esc}^3(r)-2\pi\sigma_B^2v_{esc}(r) \Bigg].
\end{eqnarray}
The number of stars in unit radial interval at $r$ is then given by 
\be N(r)=4\pi r^2 n_B(r).\ee

\begin{figure}[h]
\centering
\includegraphics[width=17cm]{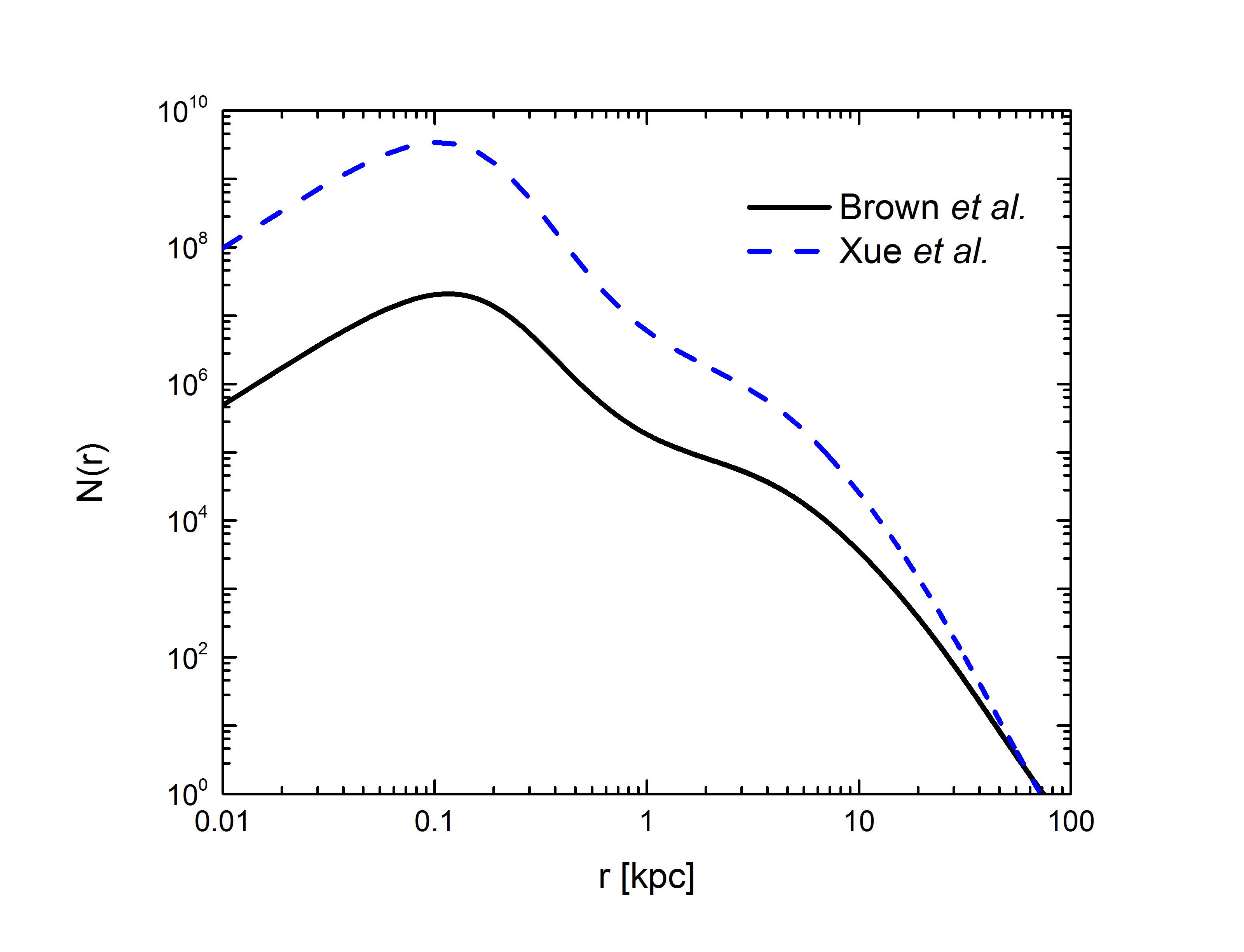}
\caption{The theoretical radial distribution of BHB and BS stars expected from the velocity distribution measured by Brown et al. and Xue et al. The distribution is shown for the $\Sigma_{d,\odot}$=55 M$_\odot$ pc$^{-2}$, $r_{tn}=3.0$ kpc disk with $\rho_{DM,c}=100$ GeV cm$^{-3}$, $\sigma_{DM}=220$ km s$^{-1}$, and $\sigma_{BHB} = \sigma_{BS} = 115$ km s$^{-1}$ for Brown et al. and $\sigma_{BHB}=106$ km s$^{-1}$ for Xue et al.}\label{NBHB}
\end{figure}

\begin{figure}[h]
\centering
\includegraphics[width=17cm]{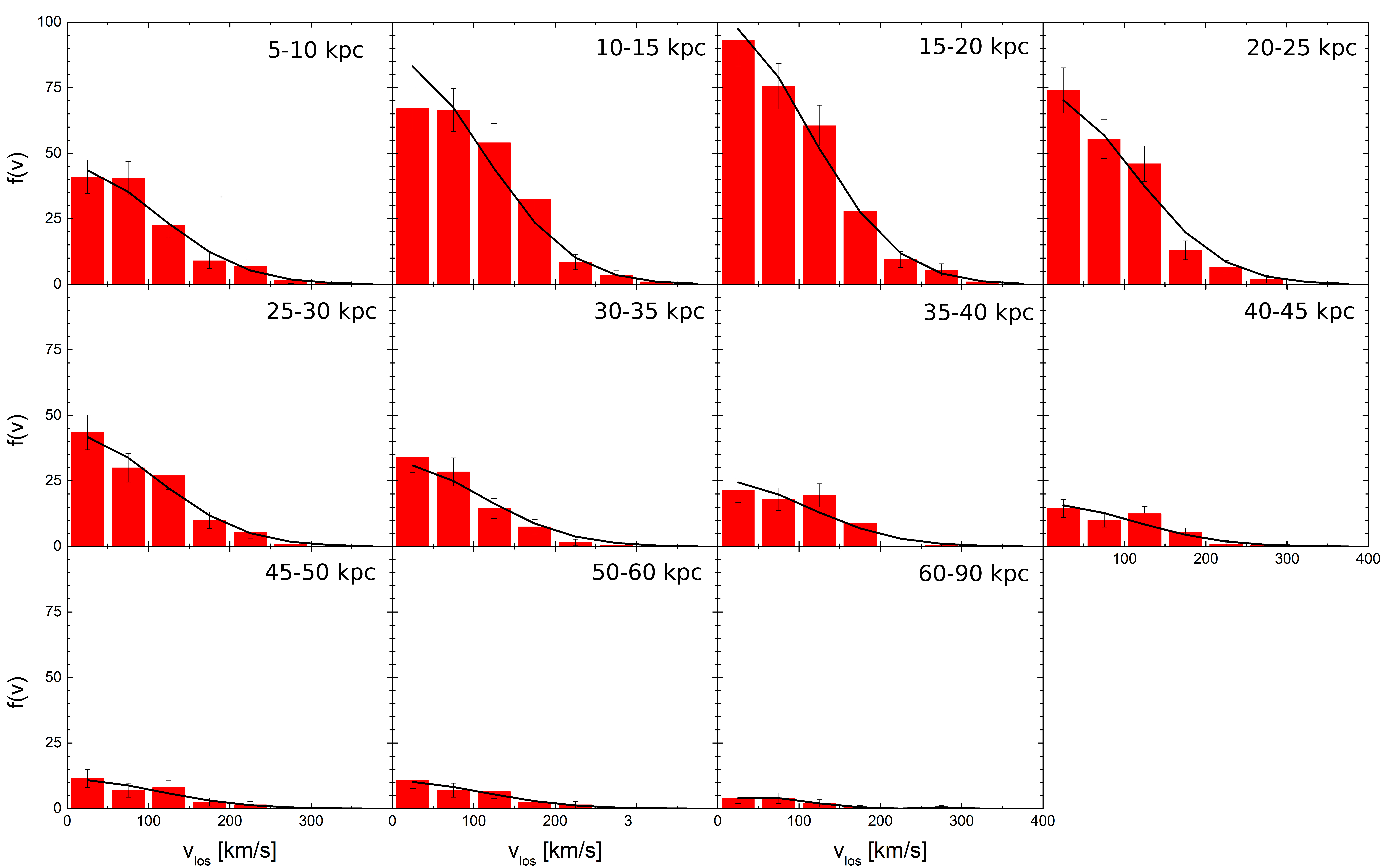}
\caption{The line-of-sight velocity distribution of BHB stars from \citet{Xue} is presented in radial velocity and radial bins in the red histogram with error bars. The distribution computed for the best-fit dark matter model for the $\Sigma_{d,\odot}$=55 M$_\odot$ pc$^{-2}$, $r_{tn}=3.0$ kpc, disk is shown as points, connected for clarity. In this model, $\rho_{DM,c}=100$ GeV cm$^{-3}$, $\sigma_{DM}=220$ km s$^{-1}$,  $\Phi_0/\sigma_{DM}^2$=10 and $\sigma_{BHB}=106$ km s$^{-1}$.}
\label{fXue40}
\end{figure}

\begin{figure}[h]
\centering
\includegraphics[width=17cm]{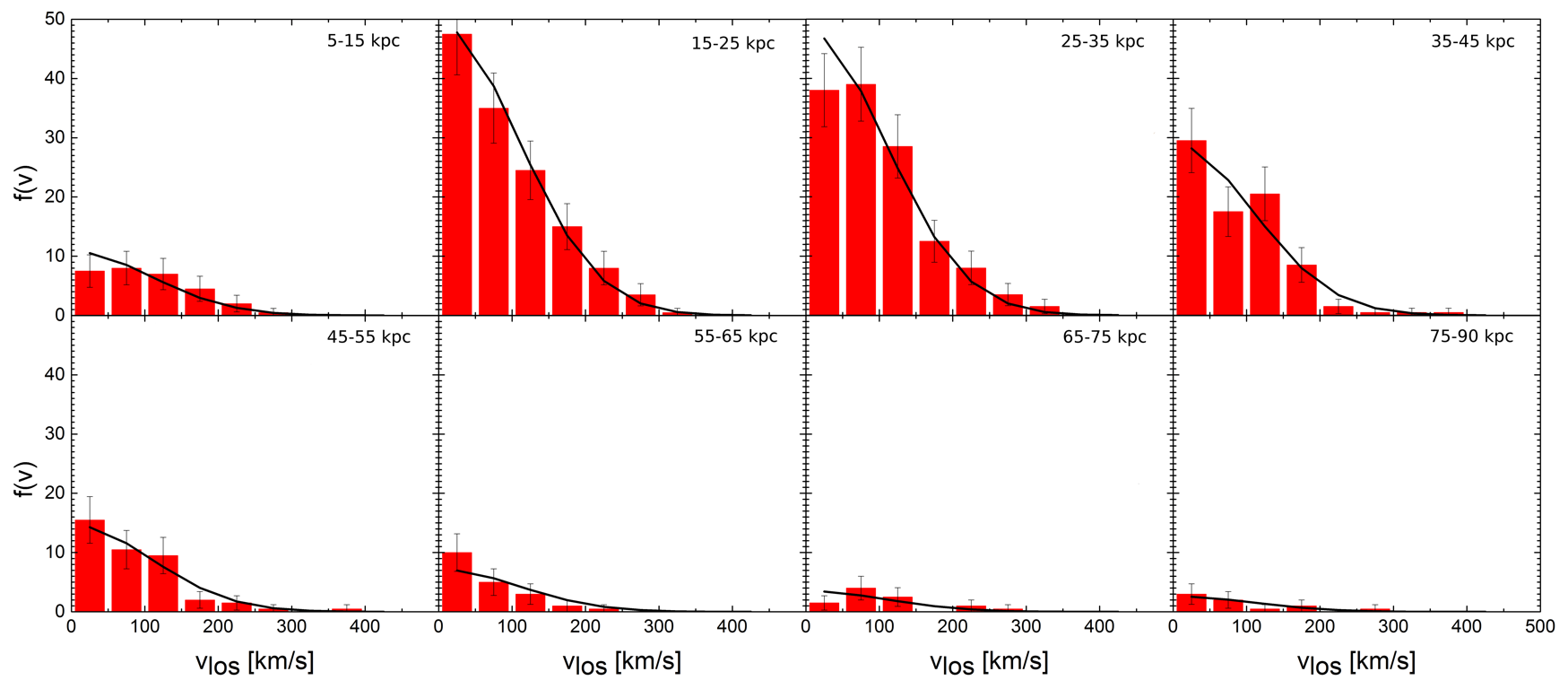}
\caption{The line-of-sight velocity distribution of BHB and BS stars from \citet{Brown} is presented in radial velocity and radial bins in the red histogram with error bars. The distribution computed for the best-fit dark matter model for the $\Sigma_{d,\odot}$=55 M$_\odot$ pc$^{-2}$, $r_{tn}=3.0$ kpc disk is shown as points, connected for clarity. In this model, $\rho_{DM,c}=200$ GeV cm$^{-3}$, $\sigma_{DM}=230$ km s$^{-1}$,  $\Phi_0/\sigma_{DM}^2$=10 and $\sigma_{BHB, BS}=115$ km s$^{-1}$.}
\label{fBrown40}
\end{figure}

The distributions of BHB and BS stars with $\sigma_B=106$ km s$^{-1}$ for the \citet{Xue} BHB stars and $\sigma_B=115$ km s$^{-1}$ for the \citet{Brown} BHB and BS stars calculated for the best fitting model with $\Sigma_{d,\odot}=55$ M$_\odot$ pc$^{-2}$ are shown in Fig. \ref{NBHB}. A single choice for the parameter $\rho_{B}$ should, in principle, fit the total number of stars observed in each radial bin when we include all the stars at that location irrespective of their velocities. Unfortunately, this does not happen, and the observed number of stars, especially at galactocentric distances below $\sim$20-30 kpc, fall short of the theoretical expectation. The most likely explanation for this is that the samples of BS and BHB stars are incomplete in these regions. In fact, the radial distribution presented by \citet{Xue} and \citet{Brown} also differ significantly from each other. We therefore integrate $4\pi r^2 F_B(r,v_r)$ over radial location and radial velocity, $\Delta r(r_i)$ and $\Delta v_r(r_i,v_j)$ respectively, and compare it with similarly binned observational data after normalizing the model estimates to the total number of stars observed in each individual bin separately. A selection of the fits are shown in Figs. \ref{fXue40} and \ref{fBrown40} for the disk model with $r_{tn}=3.0$ kpc and $\Sigma_{d,\odot}=55$ M$_\odot$.

For finding out how well the different models fit the observations, we define $f_{ij}$ as the theoretical expectation for the number of stars in the radial bin $r_i$ and the velocity bin $v_{rj}$. If $n_{ij}$ is the actual number of stars observed in this bin, then using Poisson statistics, we may define the likelihood function for each model to be 
\be L=\prod_{ij}\frac{e^{-f_{ij}}f_{ij}^{n_{ij}}}{n_{ij}!}.\label{Pstat}\ee
After the likelihoods for all the models were calculated, they were normalized to the model with the greatest likelihood, namely with $\rho_{DM,c}=100$ GeV cm$^{-3}$, $\sigma_{DM}=220$ km s$^{-1}$, $\Sigma_{d,\odot}=55$ M$_\odot$ pc$^{-2}$, and $r_{tn}=3.0$ kpc. The likelihood for each dark matter model that fits the rotation curve is show in the final rows of Tables \ref{T25}-\ref{T35c}. 

Most of the models with high probability, according to the BHB and BS analysis, occur for $\rho_{DM,c}\gtrsim100$ GeV cm$^{-3}$. Also, except for two specific cases, one in each of the $r_{tn}=2.5$ and 3.5 kpc scenarios, the models with $L>0.7$ are found for the choice $r_{tn}=3.0$ kpc. All models with $L>0.8$ are found for $r_{tn}=3.0$ kpc and $\Sigma_{d,\odot}=55$ M$_{\odot}$ pc$^{-2}$.  Most models for the $r_{tn}=3.5$ kpc disk, though they fit the rotation curve, do not predict enough stars at large velocities to agree with the BHB and BS data (Note the sharply falling rotation curves at large distances in these models.) and have a likelihood $L=0$. 

If we consider only the models with $L>0.7$, we find the following range in the parameters for the dark matter particles in the Galaxy: 
 \bd930.8 ~\textrm{km s$^{-1}$}~\lesssim~ v_{esc}(0)~\lesssim~983.9 ~\textrm{km s$^{-1}$}\ed
 \bd597.1 ~\textrm{km s$^{-1}$} ~\lesssim~ v_{esc}(R_0)~\lesssim~705.0 ~\textrm{km s$^{-1}$}\ed
\bd75 ~\textrm{GeV cm$^{-3}$}~\lesssim~\rho_{DM,c}~\lesssim~500 ~\textrm{GeV cm$^{-3}$}\ed
 \bd0.395 ~\textrm{GeV cm$^{-3}$}~\lesssim~\rho_{DM}(R_0)~\lesssim~0.758~ \textrm{GeV cm$^{-3}$}\ed 
 \bd462.5~ \textrm{km s$^{-1}$}~\lesssim~<v_{DM}^2(R_0)>^{1/2}~\lesssim~546.1~ \textrm{km s$^{-1}$}\ed
 \bd 1.60\times10^{12}~\textrm{ M}_{\odot}~\lesssim~M_{Galaxy}~\lesssim~2.00\times10^{12}~\textrm{ M}_{\odot}.\ed

\section{Discussion}
\begin{figure}[h]
\centering
\subfigure[]{
\includegraphics[width=0.48\textwidth]{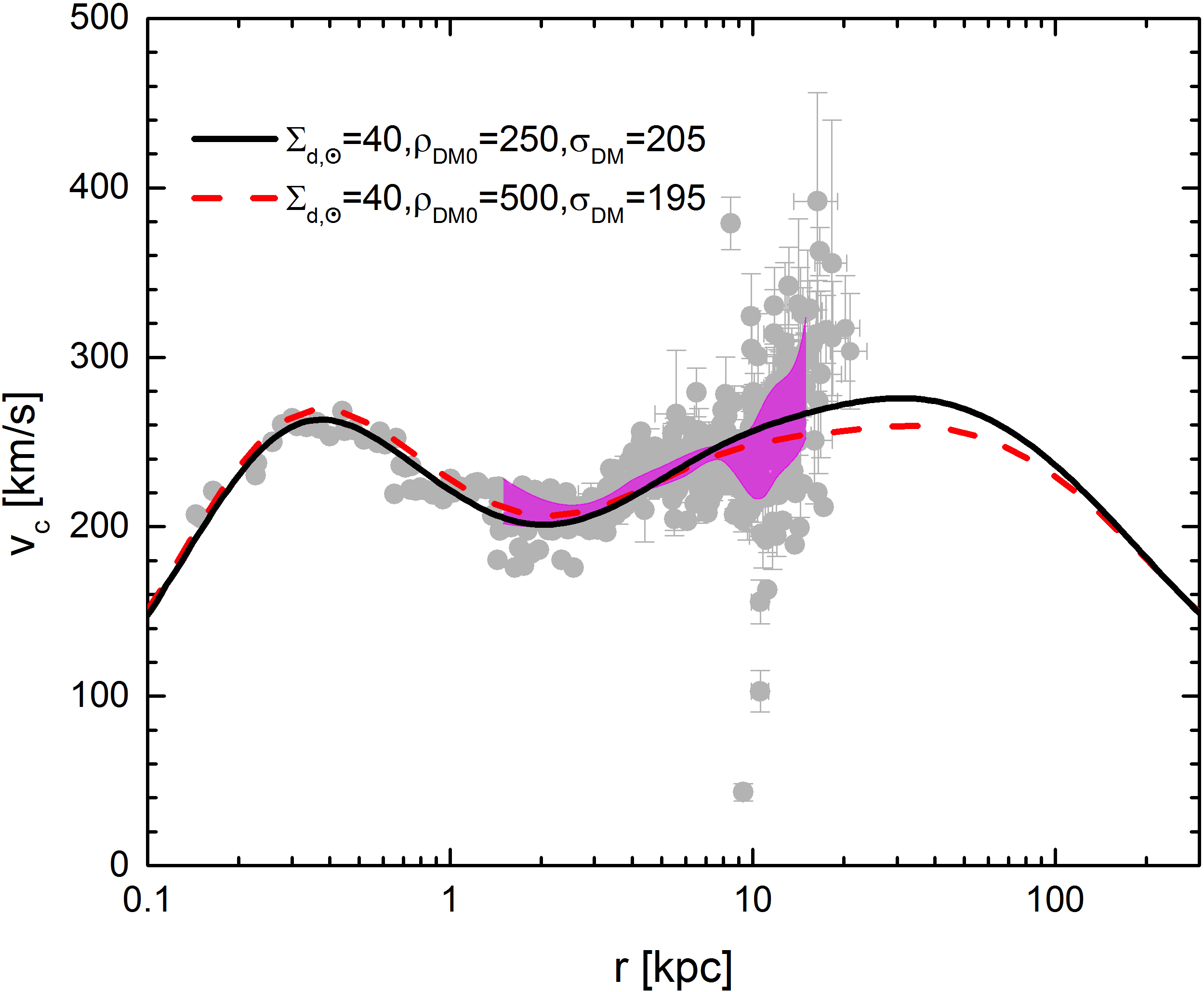}}\hfill
\subfigure[]{
\includegraphics[width=0.48\textwidth]{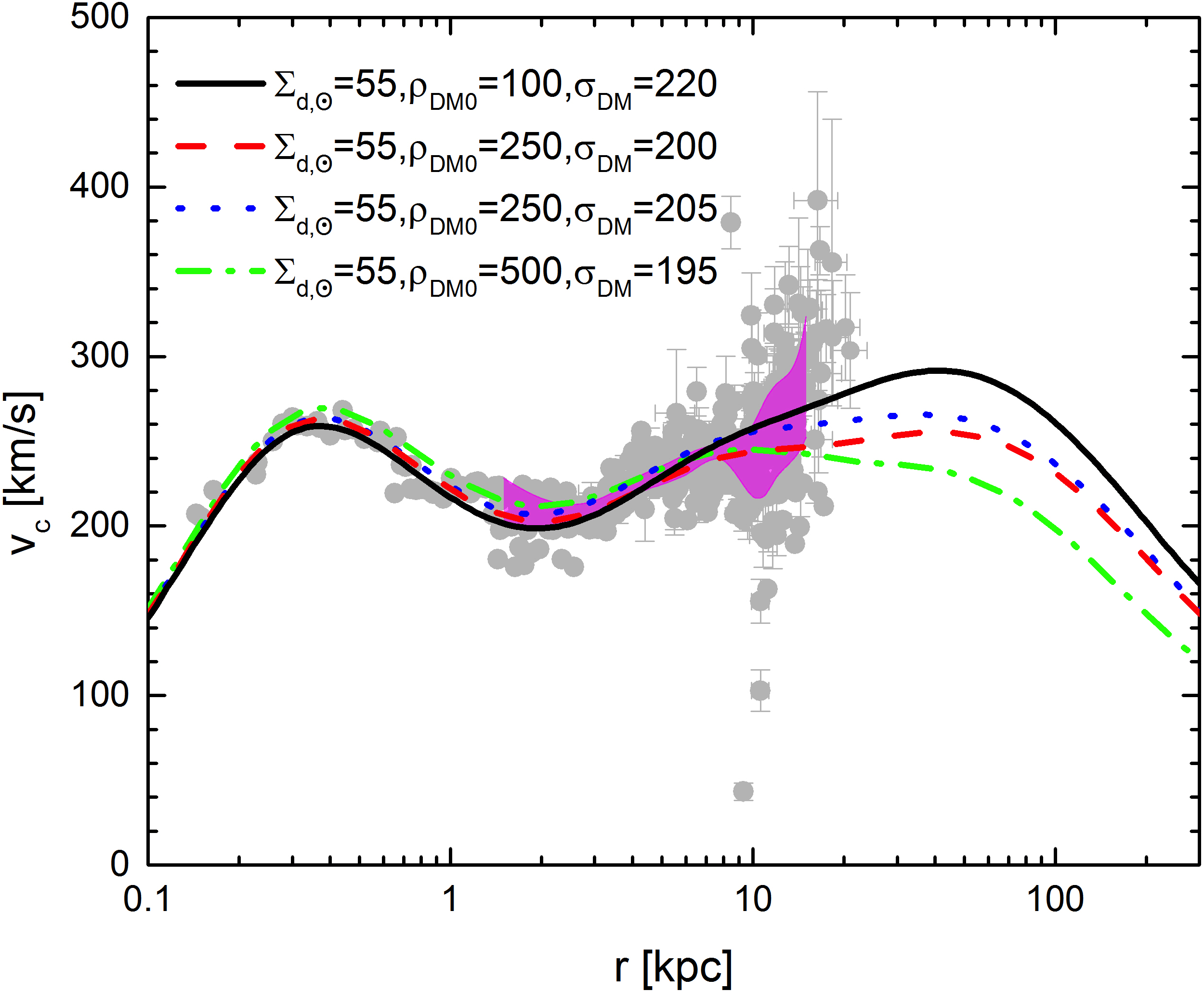}}
\subfigure[]{
\includegraphics[width=0.48\textwidth]{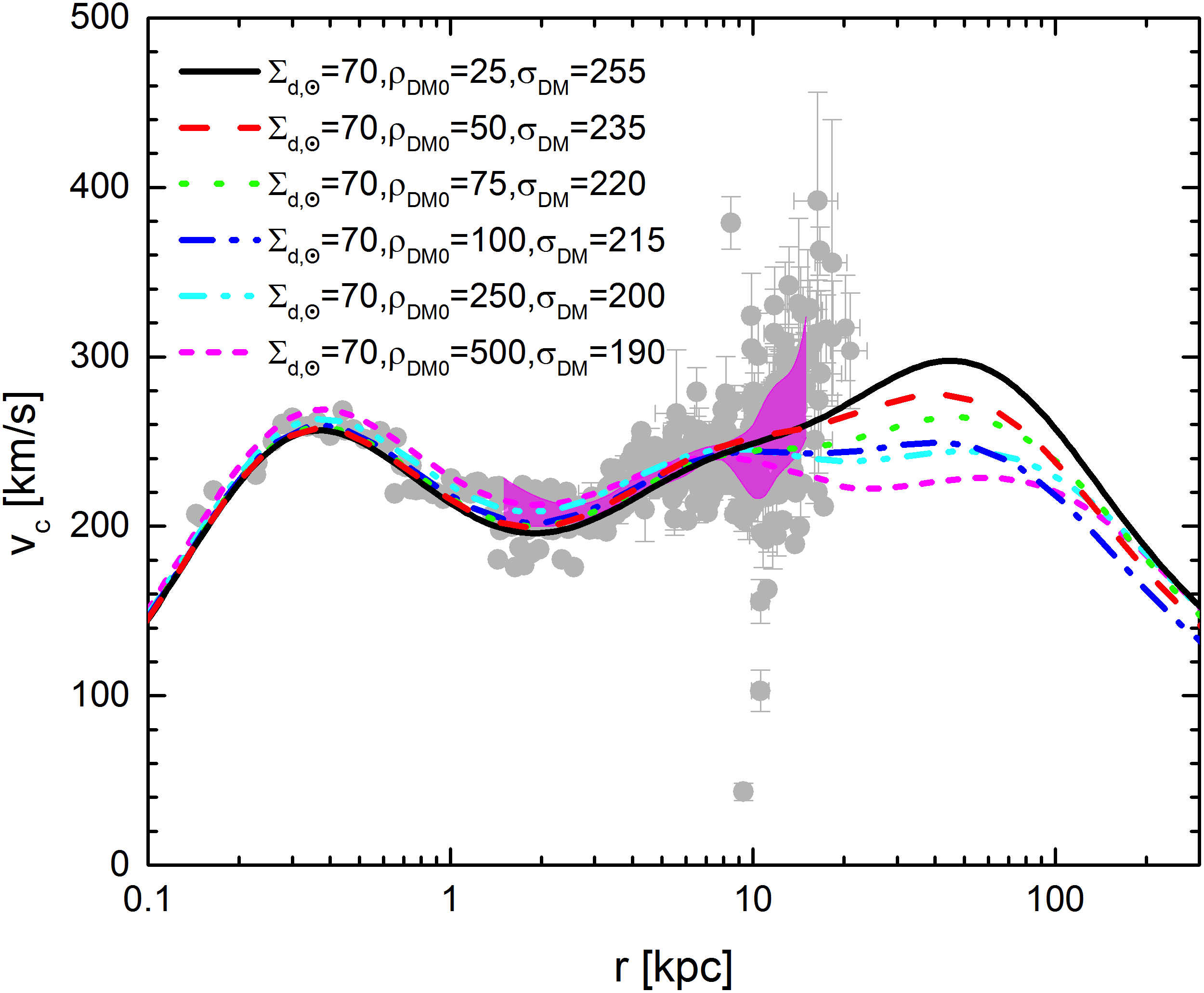}}
\caption{The rotation curves are shown for the dark matter models which pass through the band encompassing 2/3 of the observations in 1 kpc bins for $r_{tn}=3.0$ kpc and (a) $\Sigma_{d,\odot}=40$ M$_{\odot}$ pc$^{-2}$, (b) $\Sigma_{d,\odot}=55$ M$_{\odot}$ pc$^{-2}$, and (c) $\Sigma_{d,\odot}=70$ M$_{\odot}$ pc$^{-2}$}\label{RotationCurves}
\end{figure}

\begin{figure}[h]
\centering
\subfigure[]{
\includegraphics[width=0.48\textwidth]{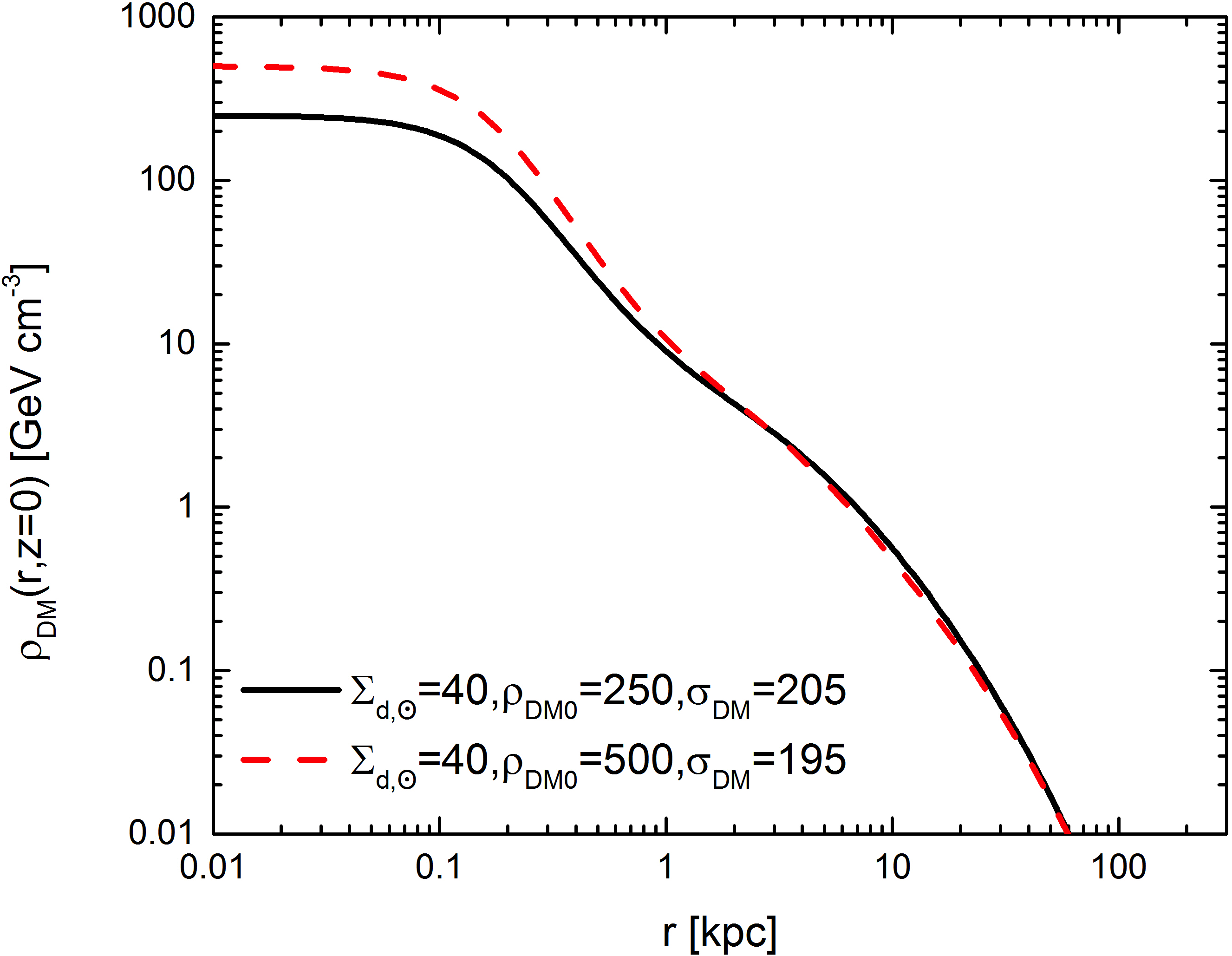}}\hfill
\subfigure[]{
\includegraphics[width=0.48\textwidth]{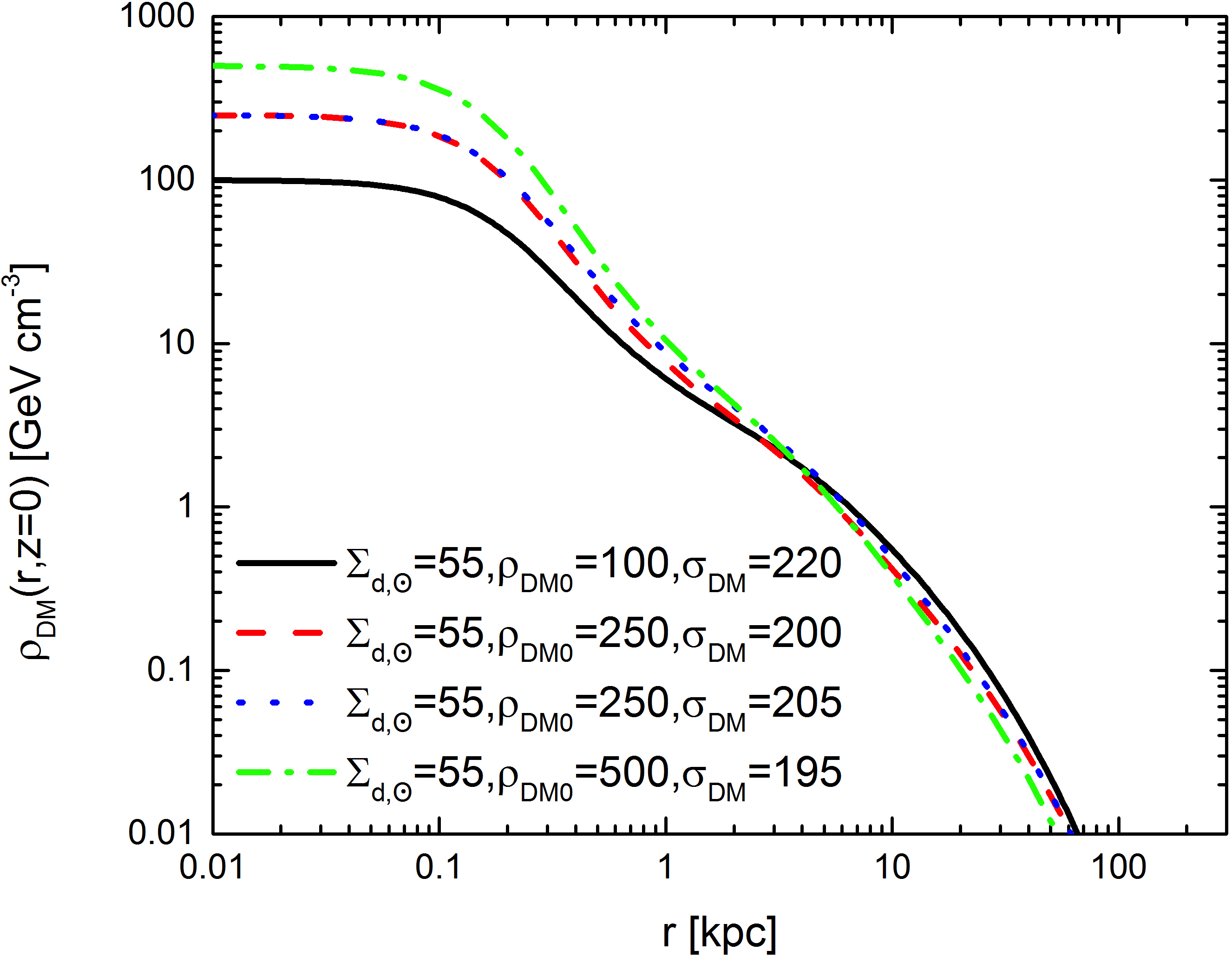}}
\subfigure[]{
\includegraphics[width=0.48\textwidth]{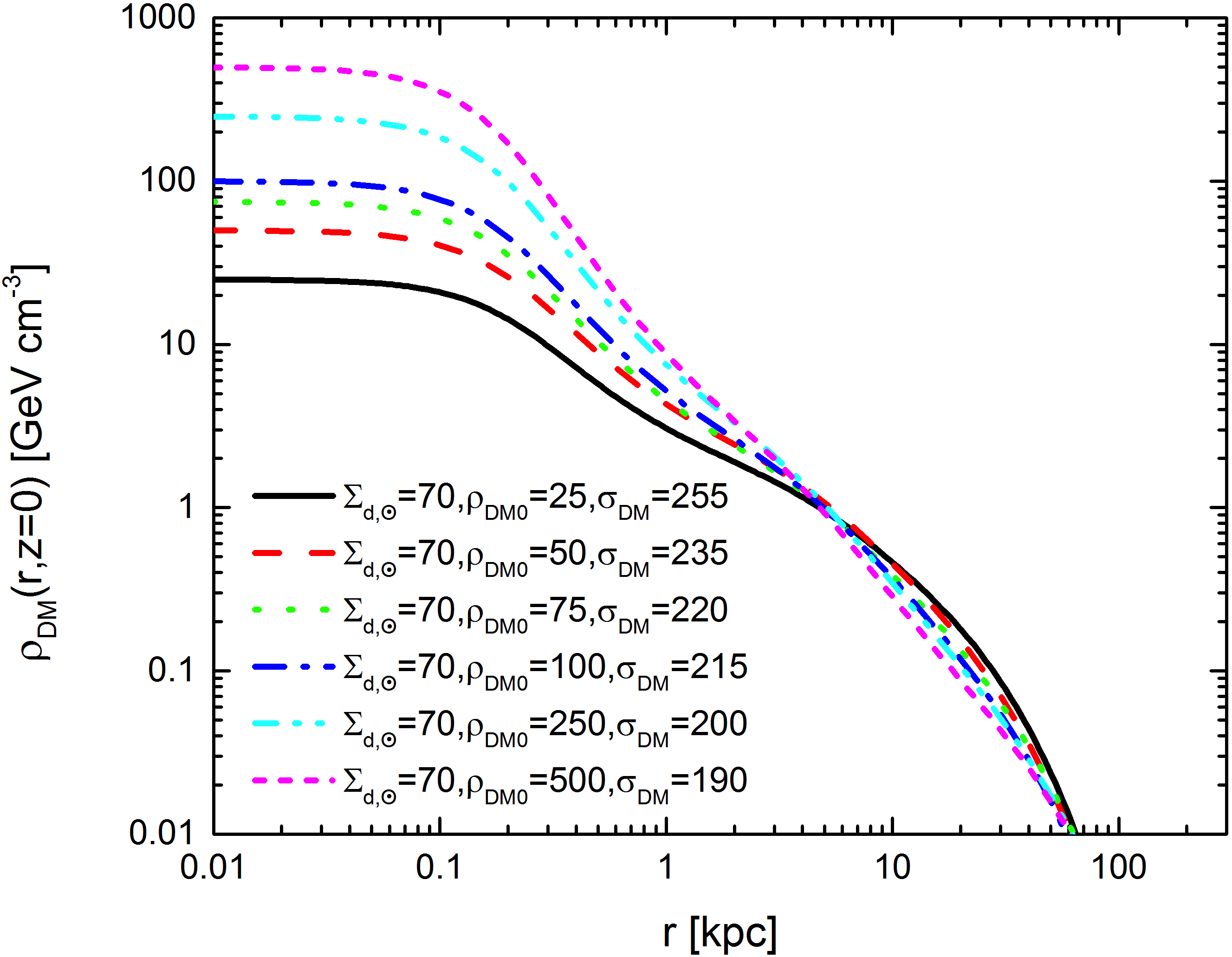}}
\caption{The density profiles for dark matter models which best fit the rotation curve are shown for $r_{tn}$ and: (a) $\Sigma_{d,\odot}=40$ M$_{\odot}$ pc$^{-2}$, (b) $\Sigma_{d,\odot}=55$ M$_{\odot}$ pc$^{-2}$, and (c) $\Sigma_{d,\odot}=70$ M$_{\odot}$ pc$^{-2}$. }\label{Densities}
\end{figure}

\begin{figure}[h]
\centering
\subfigure[]{
\includegraphics[width=0.45\textwidth]{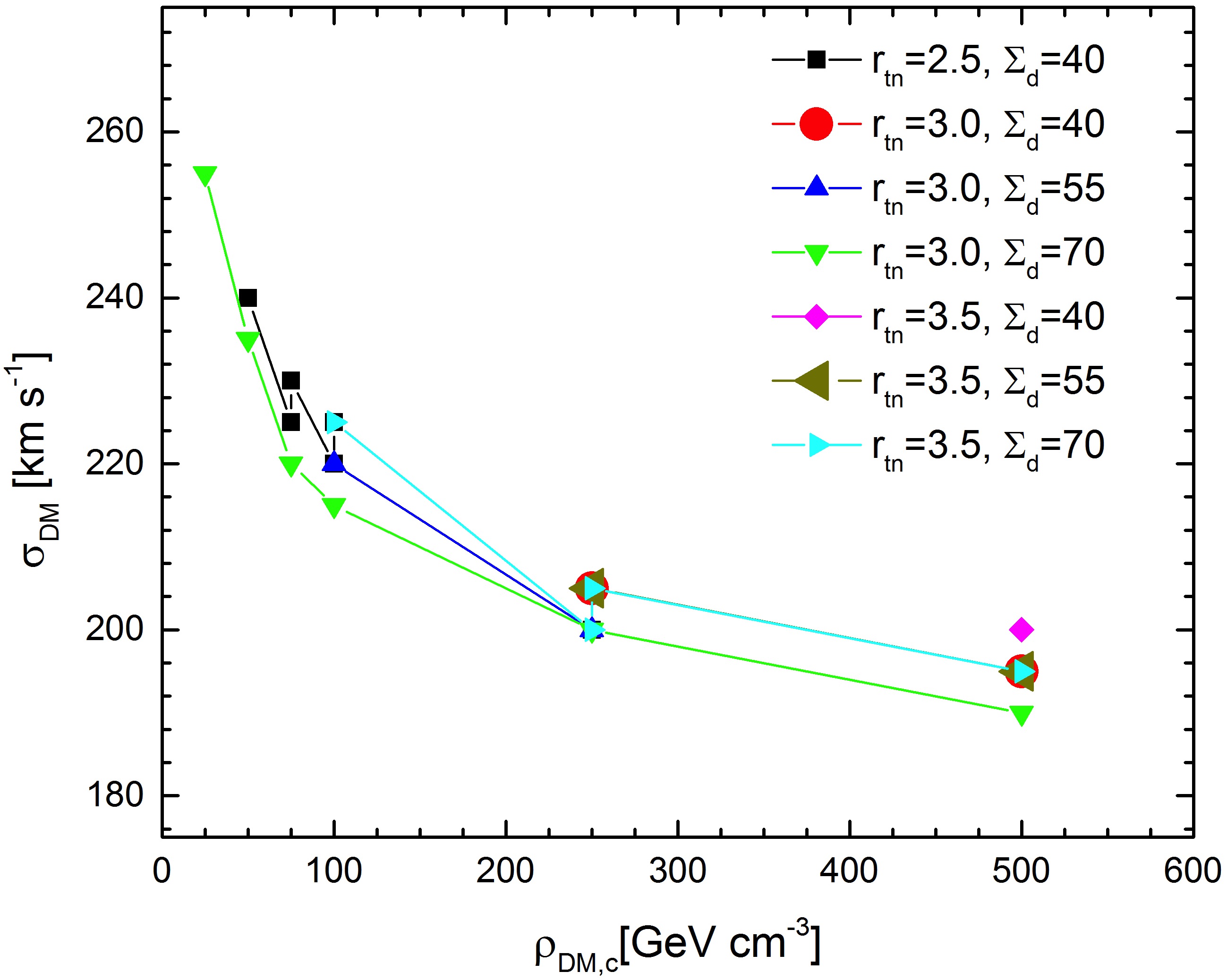}}
\subfigure[]{
\includegraphics[width=0.45\textwidth]{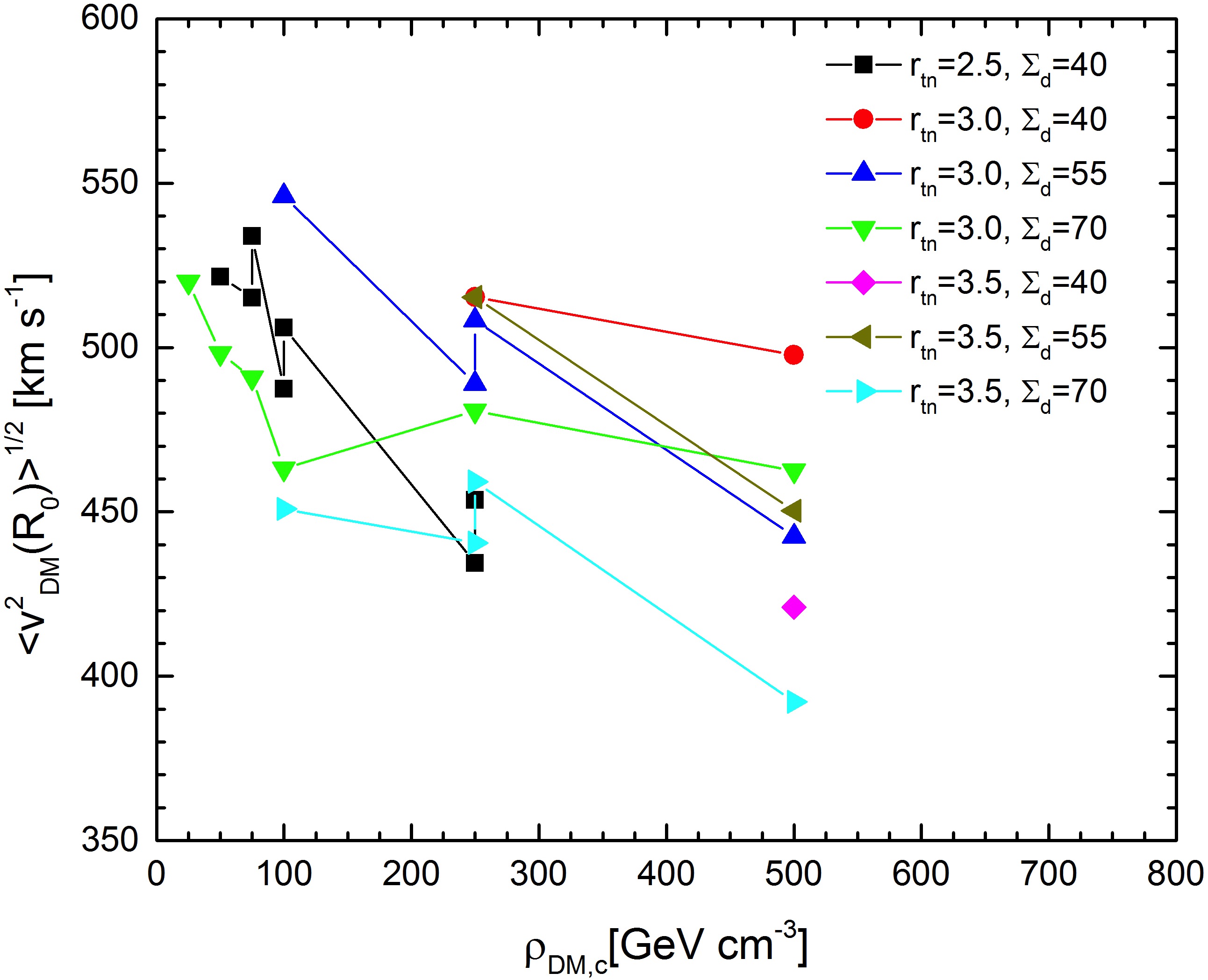}}
\subfigure[]{
\includegraphics[width=0.45\textwidth]{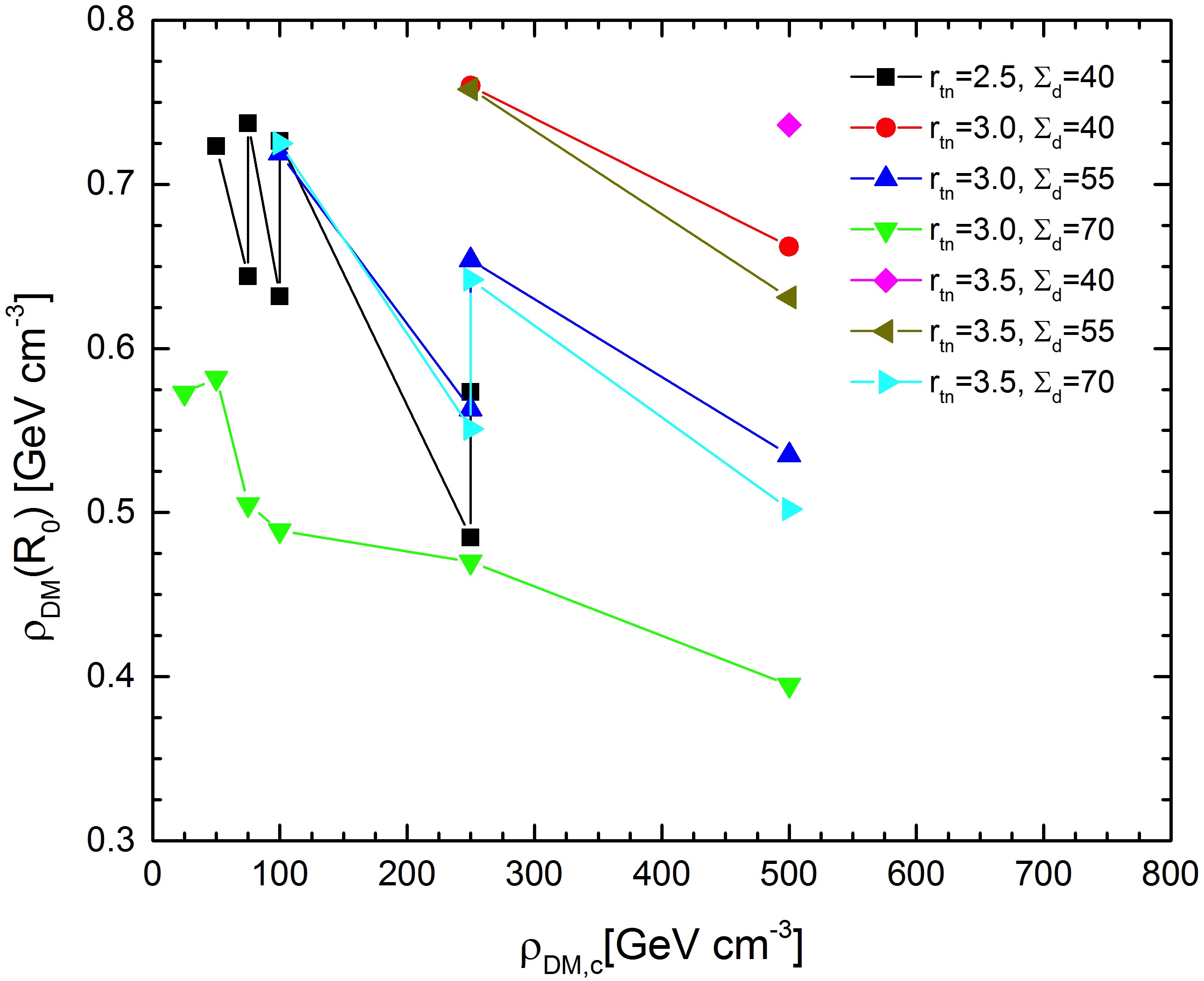}}
\subfigure[]{
\includegraphics[width=0.45\textwidth]{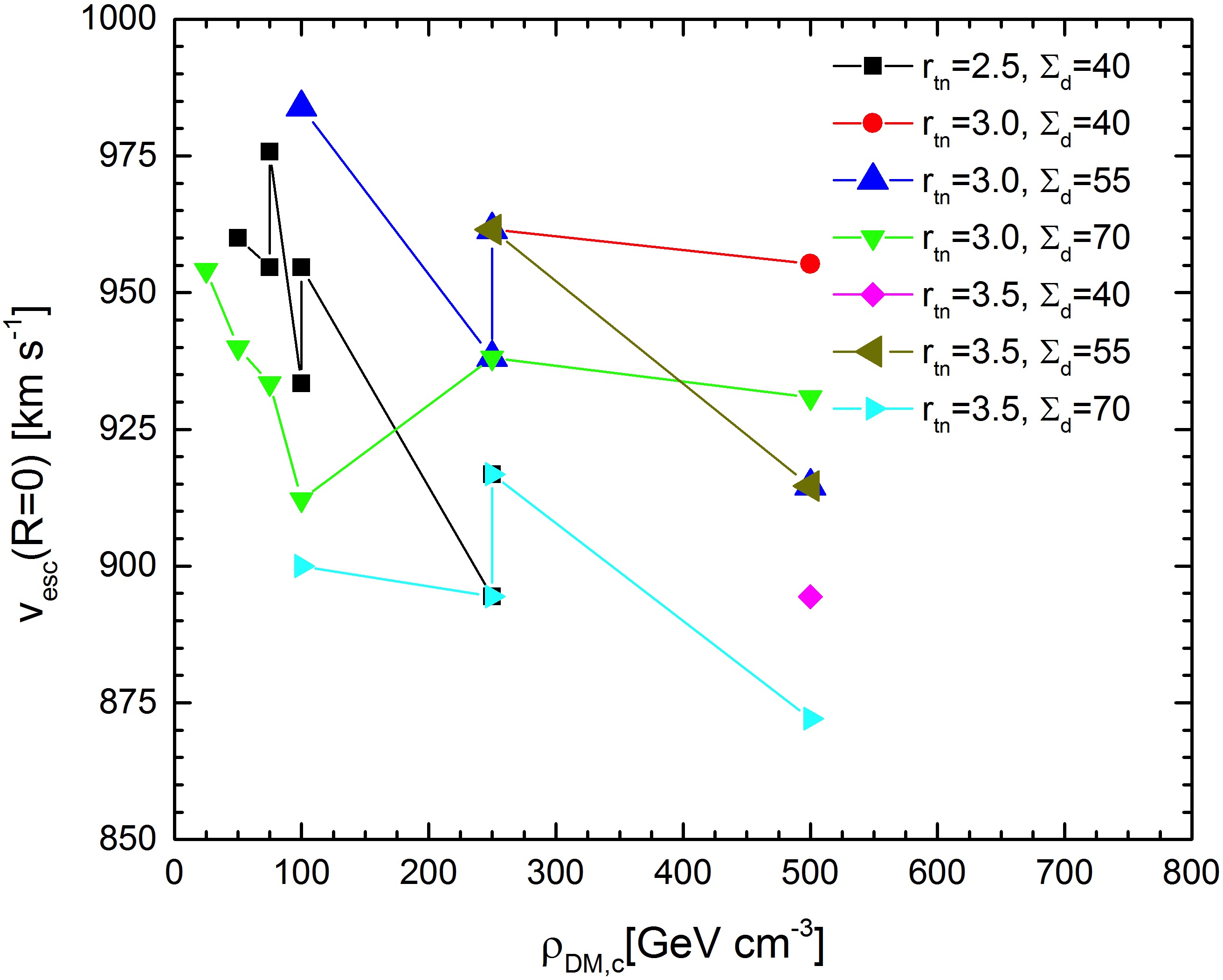}}
\subfigure[]{
\includegraphics[width=0.45\textwidth]{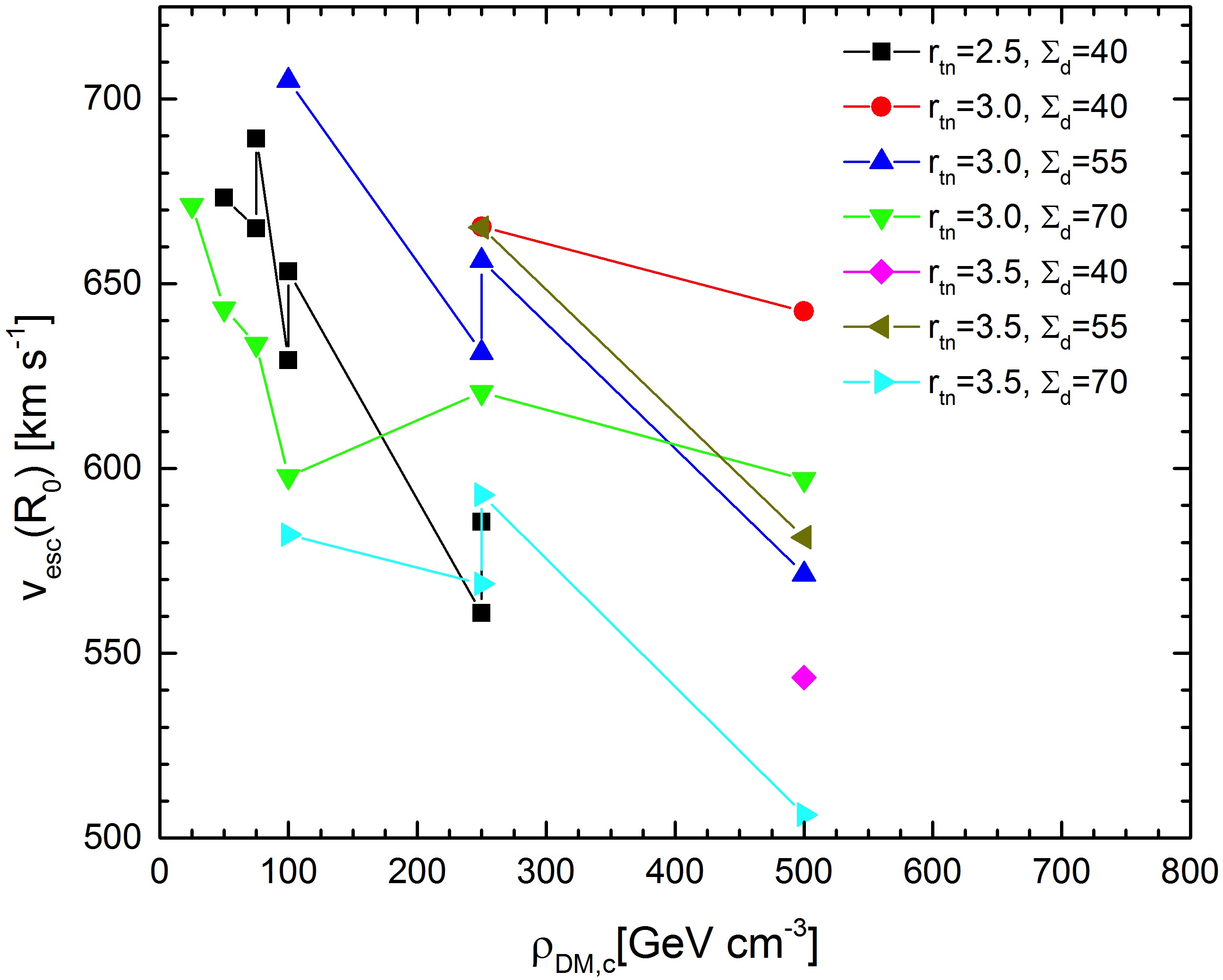}}
\subfigure[]{
\includegraphics[width=0.45\textwidth]{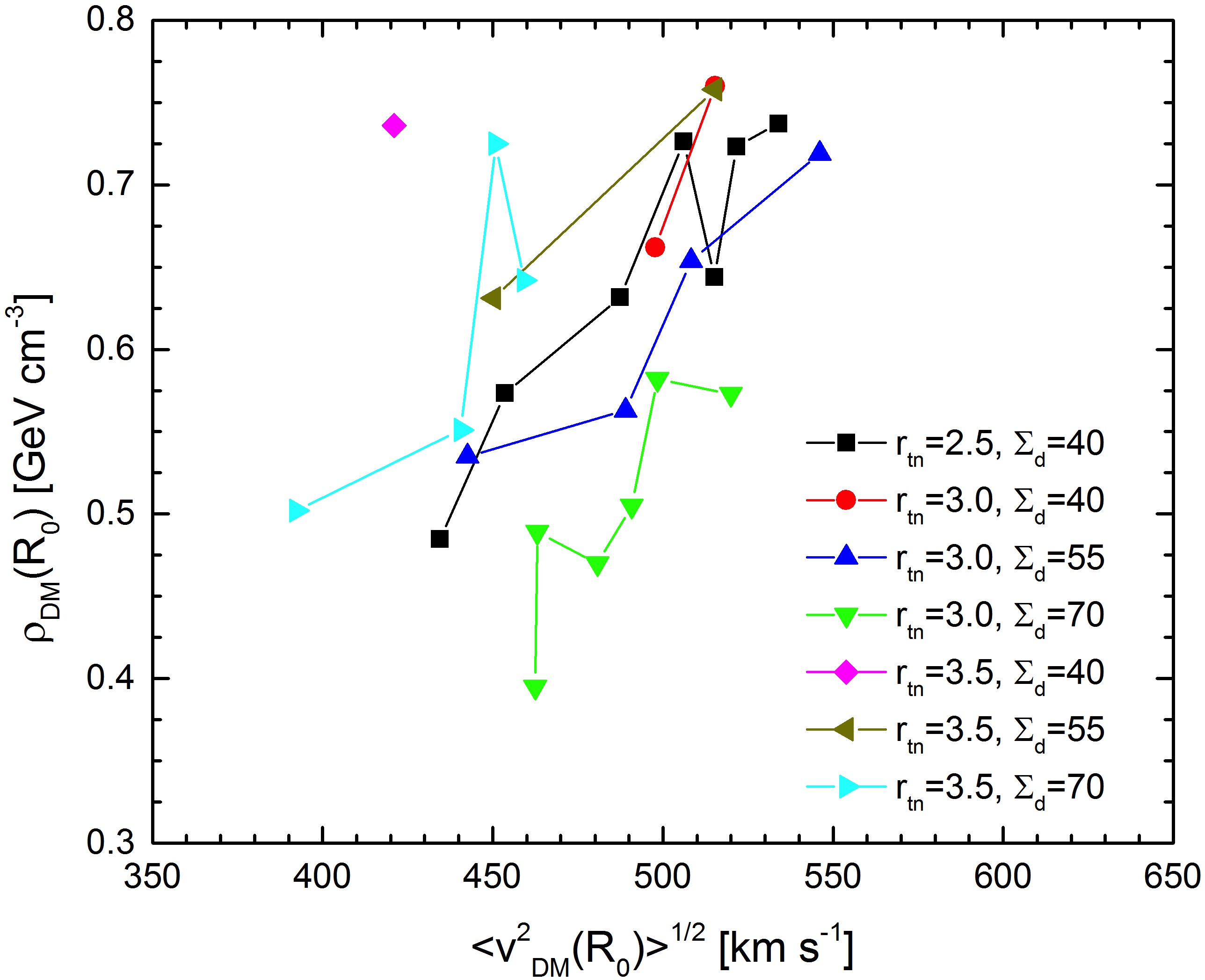}}
\caption{The correlation between the various model parameters of dark matter that fit the rotation curve of the Galaxy.}\label{CrossPlots}
\end{figure}

\begin{figure}[h]
\centering
\subfigure[]{
\includegraphics[width=0.48\textwidth]{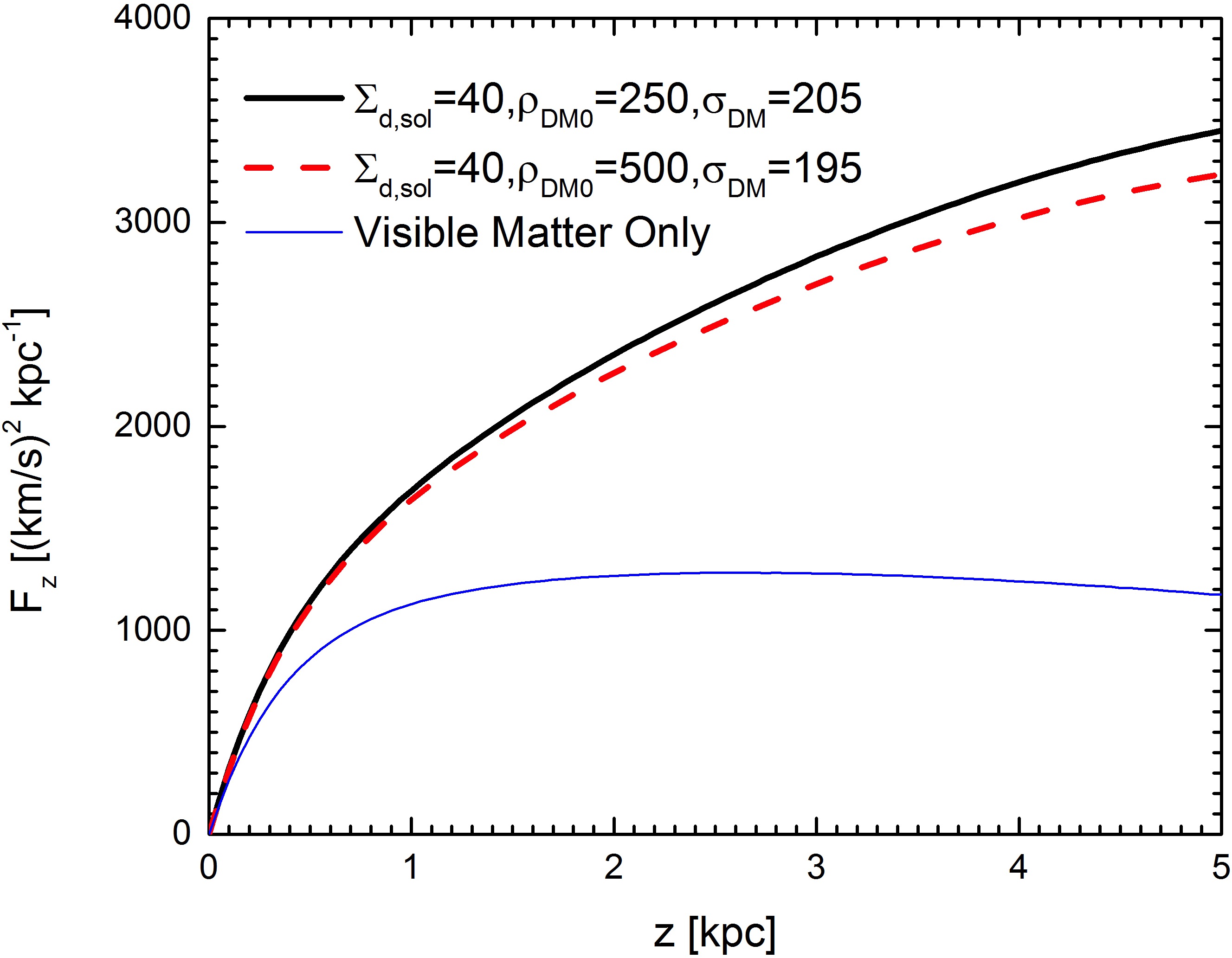}}\hfill
\subfigure[]{
\includegraphics[width=0.48\textwidth]{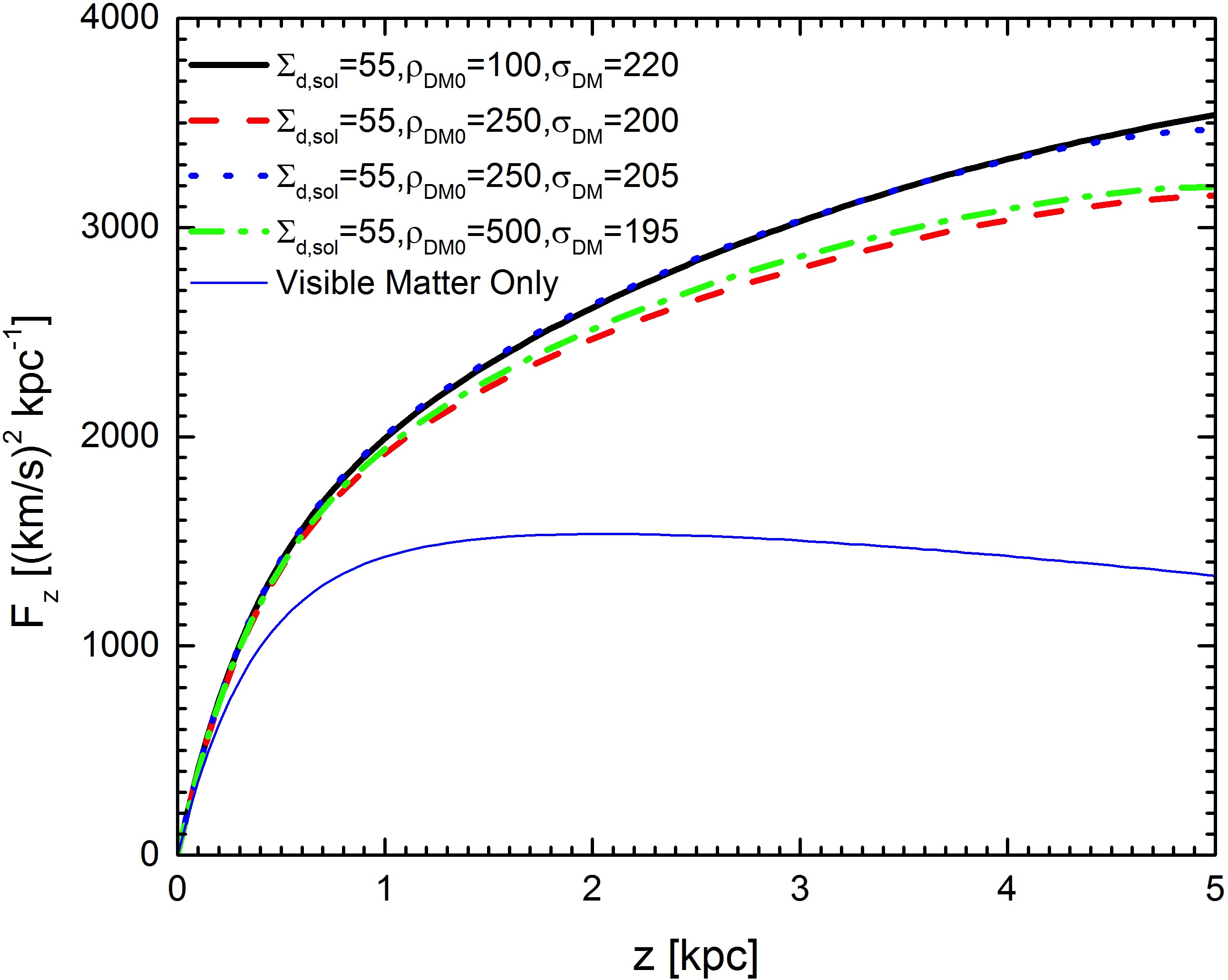}}
\subfigure[]{
\includegraphics[width=0.48\textwidth]{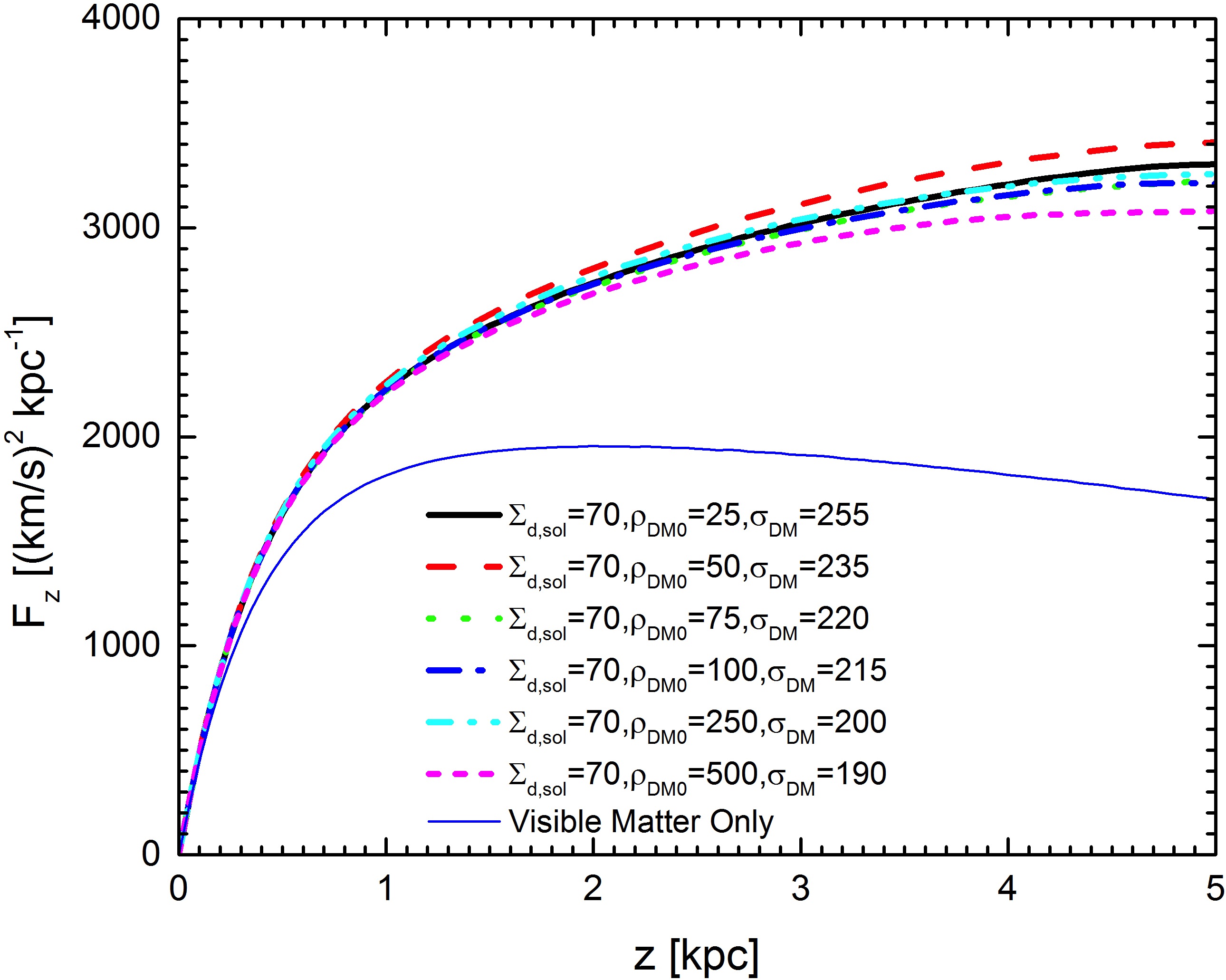}}
\caption{The vertical force profiles at the galactocentric distance of the Sun for dark matter models which best fit the rotation curve are shown for $r_{tn}=3.0$ kpc and: (a) $\Sigma_{d,\odot}=40$ M$_{\odot}$ pc$^{-2}$, (b) $\Sigma_{d,\odot}=55$ M$_{\odot}$ pc$^{-2}$, and (c) $\Sigma_{d,\odot}=70$ M$_{\odot}$ pc$^{-2}$.}\label{VerticalForce}
\end{figure}

While the observations of the rotation curve remain the best probes of the Galactic potential out to $\sim20$ kpc, the high dispersion in the observed rotation speeds, especially at large distances, does not allow  a precise determination of the parameters characterizing the phase-space distribution of Galactic dark matter. Among the models where the rotation curve is confined to lie within a narrow band encompassing 2/3 of the observed data, a wide range in the parameters is allowed, as shown in Tables \ref{T25}-\ref{T35c} and Figs. \ref{RotationCurves}-\ref{VerticalForce}. It may be possible to further constrain the dark matter phase-space distribution by comparing the vertical force exerted by the Galactic potential at $R_0$ displayed for several models in Fig. \ref{VerticalForce} with the stellar observations perpendicular to the plane in the Solar neighborhood as an extension of the analysis by \citet{Kuijken89b}. 

The model providing the best fit for disks with surface densities within observational constraints to both the rotation curve and the BHB/BS distributions occurs for the following dark matter phase-space distribution parameters: $\rho_{DM,c}$=100 GeV cm$^{-3}$, $\Psi_0/\sigma_{DM}^2$=10, and $\sigma_{DM}=220$ km s$^{-1}$. The corresponding properties of the dark matter relevant to its direct detection are $<v^2(R_0)_{DM}>^{1/2}\approx546 $ km s$^{-1}$ and $\rho_{DM}(R_0)\approx0.72$ GeV cm$^{-3}$. This local dark matter density is notably higher than the currently adopted Standard Halo Model. The effect this increase has on the expected rates of direct and indirect detection predictions will be discussed in a companion paper.

Better observations of the distance of the Sun from the Galactic center and the value of the rotation speed at the Solar circle as well as observations of the rotation curve beyond $R_0$, especially at distances beyond 20 kpc, would allow for better constraints on the phase-space distribution of dark matter. The currently observed distributions of BHB and BS stars have large uncertainties as to their absolute spatial densities and do not place tight bounds on the dark matter phase-space distribution on their own. This was shown by the wide range of dark matter models that adequately reproduced the velocity distributions at various distances compiled by Xue et al. and Brown et al. The wide range of allowed parameters shows that more precise astronomical observations, especially pertaining to the surface mass density of the disk and its radial scale length, are needed to narrow down the parameters of  the phase-space distribution of dark matter  and to be able to correctly interpret the results of direct and indirect dark matter detection experiments. As far as the present status of the observations are concerned, assuming the surface density of the Galactic disk is $\sim55$ M$_\odot$ pc$^{-2}$, a value in the middle of the observational constraints, and a thin disk scale length of 3.0 kpc, the best choice for the parameters of Galactic dark matter (L>0.7) are the following: $\rho_{DM,c}\approx100-250$ GeV cm$^{-3}$, $\rho_{DM}(R_0)\approx0.56-0.72$ GeV cm$^{-3}$, and $<v_{DM}^2(R_0)>^{1/2}\approx$490-550 km s$^{-1}$, which yield a total mass of the Galaxy including the dark matter halo of $M_{Galaxy}\approx1.68-2.00\times10^{12}$ M$_{\odot}$.

\begin{table}[tp]
\caption{Models for the $r_{tn}=2.5$ and $\Sigma_{d,\odot}$=40 M$_\odot$ pc$^{-2}$ disk.}\label{T25}
\begin{centering}
\begin{tabular}{lcccccc}
\tableline\tableline %
$\rho_{DM,c}$ [GeV cm$^{-3}$]&25&50&75&100&250&500\\
$\sigma_{DM}$ [km s$^{-1}$]&--&240&225&220&205&--\\
&-- & --&230&225&--&--\\
$\Psi_0/\sigma_{DM}^2$&--&8&9&9&10&--\\
&-- &-- &9&9&--&--\\
\tableline
$v_{esc}(0)$ [km s$^{-1}$]&--&960&955&933&917&--\\
&-- &-- &976&955&--&--\\
$v_c(R_0)$ [km s$^{-1}$]&--&246&241&242&243&--\\
& --& --&250&252&--&--\\
$<v_{DM}^2(R_0)>^{1/2}$ [km s$^{-1}$]&--&522&515&487&454&--\\
&-- &-- &534&506&--&--\\
$\rho_{DM}(R_0)$ [GeV cm$^{-3}$]&--&0.723&0.644&0.632&0.573&--\\
&-- &-- &0.737&0.726&--&--\\
$v_{esc}(R_0)$ [km s$^{-1}$]&--&673&665&629&586&--\\
&-- &-- &689&653&--&--\\
$M_{galaxy}$ [10$^{12}$ M$_{\odot}$]&--&1.54&1.64&1.34&1.10&--\\
& --& --&1.71&1.40&--&--\\
\tableline
$L$ &--&0.274&0.652&0.093&0&--\\
&-- &-- &0.746&0.173&--&--\\
\tableline\tableline
\end{tabular}\\
\end{centering}
\scriptsize
In the above and following tables, the dark matter at the center of the Galaxy, $\rho_{DM,c}$, and the parameters $\sigma_{DM}$ and $\Psi_0/\sigma_{DM}^2$ are the free parameters for the lowered-isothermal model of the phase-space distribution of Galactic dark matter, as explained in Section 4. We also show the following derived quantities:  the escape speed from the Galactic center, $v_{esc}(0)$, the rotation speed at the Solar location, $v_c(R_0)$, the root-mean-square velocity of dark matter particles at the Solar location, $<v_{DM}^2(R_0)>^{1/2}$, the local dark matter density,  $\rho_{DM}(R_0)$, the escape speed from the Solar location, $v_{esc}(R_0)$, and the total mass of the Galaxy including both visible and dark matter, $M_{galaxy}$. Also shown is the Poisson likelihood $L$ (eq. \ref{Pstat}) computed by comparing predictions from the dark matter models to the radial and velocity distributions of BHB and BS stars, as explained in Section 5.2. For a given value of $\rho_{DM,c}$, entries with two values of $\sigma_{DM}$ indicate instances where both values of $\sigma_{DM}$ had corresponding rotation curves that were consistent with 2/3 of the available observations. We show the derived quantities for both values of  $\sigma_{DM}$. \end{table}

\begin{table}[tp]
\centering
\caption{Models for the $r_{tn}=3.0$ and $\Sigma_{d,\odot}$=40 M$_\odot$ pc$^{-2}$ disk.}\label{T30a}
\begin{tabular}{lcccccc}
\tableline\tableline %
$\rho_{DM,c}$ [GeV cm$^{-3}$]&25&50&75&100&250&500\\
$\sigma_{DM}$ [km s$^{-1}$]&--&--&--&--&205&195\\
$\Psi_0/\sigma_{DM}^2$&--&--&--&--&11&12\\
\tableline
$v_{esc}(0)$ [km s$^{-1}$]&--&--&--&--&962&955\\
$v_c(R_0)$ [km s$^{-1}$]&--&--&--&--&250&244\\
$<v_{DM}^2(R_0)>^{1/2}$ [km s$^{-1}$]&--&--&--&--&515&498\\
$\rho_{DM}(R_0)$ [GeV cm$^{-3}$]&--&--&--&--&0.760&0.662\\
$v_{esc}(R_0)$ [km s$^{-1}$]&--&--&--&--&665&643\\
$M_{galaxy}$ [10$^{12}$ M$_{\odot}$]&--&--&--&--&1.61&1.62\\
\tableline
$L$ &--&--&--&--&0.777&0.791\\
\tableline\tableline
\end{tabular}\\
\end{table}

\begin{table}[tp]
\centering
\caption{Models for the $r_{tn}=3.0$ and $\Sigma_{d,\odot}$=55 M$_\odot$ pc$^{-2}$ disk.}\label{T30b}
\begin{tabular}{lcccccc}
\tableline\tableline %
$\rho_{DM,c}$ [GeV cm$^{-3}$]&25&50&75&100&250&500\\
$\sigma_{DM}$ [km s$^{-1}$]&--&--&--&220&200&195\\
&--&--&--&--&205&--\\
$\Psi_0/\sigma_{DM}^2$&--&--&--&10&11&11\\
&--&--&--&--&11&--\\
\tableline
$v_{esc}(0)$ [km s$^{-1}$]&--&--&--&984&938&915\\
&--&--&--&--&962&--\\
$v_c(R_0)$ [km s$^{-1}$]&--&--&--&251&241&245\\
&--&--&--&--&252&--\\
$<v_{DM}^2(R_0)>^{1/2}$ [km s$^{-1}$]&--&--&--&546&489&443\\
&--&--&--&--&508&--\\
$\rho_{DM}(R_0)$ [GeV cm$^{-3}$]&--&--&--&0.719&0.563&0.535\\
&--&--&--&--&0.654&--\\
$v_{esc}(R_0)$ [km s$^{-1}$]&--&--&--&705&631&571\\
&--&--&--&--&656&--\\
$M_{galaxy}$ [10$^{12}$ M$_{\odot}$]&--&--&--&2.00&1.60&1.06\\
&--&--&--&--&1.68&--\\
\tableline
$L$ &--&--&--&1&0.732&0\\
&--&--&--&--&0.834&--\\
\tableline\tableline
\end{tabular}\\
\end{table}

\begin{table}[tp]
\centering
\caption{Models for the $r_{tn}=3.0$ and $\Sigma_{d,\odot}$=70 M$_\odot$ pc$^{-2}$ disk.}\label{T30c}
\begin{tabular}{lcccccc}
\tableline\tableline %
$\rho_{DM,c}$ [GeV cm$^{-3}$]&25&50&75&100&250&500\\
$\sigma_{DM}$ [km s$^{-1}$]&255&235&220&215&200&190\\
$\Psi_0/\sigma_{DM}^2$&7&8&9&9&11&12\\
\tableline
$v_{esc}(0)$ [km s$^{-1}$]&954&940&933&912&938&931\\
$v_c(R_0)$ [km s$^{-1}$]&244&248&243&243&246&241\\
$<v_{DM}^2(R_0)>^{1/2}$ [km s$^{-1}$]&520&498&491&463&481&463\\
$\rho_{DM}(R_0)$ [GeV cm$^{-3}$]&0.573&0.582&0.505&0.489&0.470&0.395\\
$v_{esc}(R_0)$ [km s$^{-1}$]&671&643&634&598&621&597\\
$M_{galaxy}$ [10$^{12}$ M$_{\odot}$]&1.69&1.46&1.58&1.27&1.67&1.69\\
\tableline
$L$ &0.524&0.250&0.597&0.056&0.784&0.764\\
\tableline\tableline
\end{tabular}\\
\end{table}

\begin{table}[tp]
\centering
\caption{Models for the $r_{tn}=3.5$ and $\Sigma_{d,\odot}$=40 M$_\odot$ pc$^{-2}$ disk.}\label{T35a}
\begin{tabular}{lcccccc}
\tableline\tableline %
$\rho_{DM,c}$ [GeV cm$^{-3}$]&25&50&75&100&250&500\\
$\sigma_{DM}$ [km s$^{-1}$]&--&--&--&--&--&200\\
$\Psi_0/\sigma_{DM}^2$&--&--&--&--&--&10\\
\tableline
$v_{esc}(0)$ [km s$^{-1}$]&--&--&--&--&--&894\\
$v_c(R_0)$ [km s$^{-1}$]&--&--&--&--&--&251\\
$<v_{DM}^2(R_0)>^{1/2}$ [km s$^{-1}$]&--&--&--&--&--&421\\
$\rho_{DM}(R_0)$ [GeV cm$^{-3}$]&--&--&--&--&--&0.736\\
$v_{esc}(R_0)$ [km s$^{-1}$]&--&--&--&--&--&544\\
$M_{galaxy}$ [10$^{12}$ M$_{\odot}$]&--&--&--&--&--&0.67\\
\tableline
$L$ &--&--&--&--&--&0\\
\tableline\tableline
\end{tabular}\\
\end{table}

\begin{table}[tp]
\centering
\caption{Models for the $r_{tn}=3.5$ and $\Sigma_{d,\odot}$=55 M$_\odot$ pc$^{-2}$ disk.}\label{T35b}
\begin{tabular}{lcccccc}
\tableline\tableline %
$\rho_{DM,c}$ [GeV cm$^{-3}$]&25&50&75&100&250&500\\
$\sigma_{DM}$ [km s$^{-1}$]&--&--&--&--&205&195\\
$\Psi_0/\sigma_{DM}^2$&--&--&--&--&11&11\\
\tableline
$v_{esc}(0)$ [km s$^{-1}$]&--&--&--&--&962&915\\
$v_c(R_0)$ [km s$^{-1}$]&--&--&--&--&254&246\\
$<v_{DM}^2(R_0)>^{1/2}$ [km s$^{-1}$]&--&--&--&--&515&450\\
$\rho_{DM}(R_0)$ [GeV cm$^{-3}$]&--&--&--&--&0.758&0.631\\
$v_{esc}(R_0)$ [km s$^{-1}$]&--&--&--&--&665&581\\
$M_{galaxy}$ [10$^{12}$ M$_{\odot}$]&--&--&--&--&1.61&1.02\\
\tableline
$L$ &--&--&--&--&0.781&0\\
\tableline\tableline
\end{tabular}\\
\end{table}

\begin{table}[tp]
\centering
\caption{Models for the $r_{tn}=3.5$ and $\Sigma_{d,\odot}$=70 M$_\odot$ pc$^{-2}$ disk.}\label{T35c}
\begin{tabular}{lcccccc}
\tableline\tableline %
$\rho_{DM,c}$ [GeV cm$^{-3}$]&25&50&75&100&250&500\\
$\sigma_{DM}$ [km s$^{-1}$]&--&--&--&225&200&195\\
&--&--&--&--&205&--\\
$\Psi_0/\sigma_{DM}^2$&--&--&--&8&10&10\\
&--&--&--&--&10&--\\
\tableline
$v_{esc}(0)$ [km s$^{-1}$]&--&--&--&900&894&872\\
&--&--&--&--&917&--\\
$v_c(R_0)$ [km s$^{-1}$]&--&--&--&256&243&245\\
&--&--&--&--&254&--\\
$<v_{DM}^2(R_0)>^{1/2}$ [km s$^{-1}$]&--&--&--&451&441&392\\
&--&--&--&--&459&--\\
$\rho_{DM}(R_0)$ [GeV cm$^{-3}$]&--&--&--&0.725&0.551&0.502\\
&--&--&--&--&0.642&--\\
$v_{esc}(R_0)$ [km s$^{-1}$]&--&--&--&582&569&506\\
&--&--&--&--&593&--\\
$M_{galaxy}$ [10$^{12}$ M$_{\odot}$]&--&--&--&0.84&0.98&1.04\\
&--&--&--&--&0.61&--\\
\tableline
$L$ &--&--&--&0&0&0\\
&--&--&--&--&0&--\\
\tableline\tableline
\end{tabular}\\
\end{table}

\end{document}